\documentclass[pdftex,12pt]{JHEP3}
\usepackage{amsmath,amsfonts,amsbsy,latexsym,amssymb,amscd,amstext}
\usepackage[pdftex]{epsfig}

\pdfoutput=1

\def\be{\begin{equation}}
\def\ee{\end{equation}}
\def\bea{\begin{eqnarray}}
\def\eea{\end{eqnarray}}
\def\pd{\partial}
\def\a{\alpha}
\def\b{\beta}
\def\g{\gamma}
\def\d{\delta}
\def\m{\mu}
\def\n{\nu}
\def\t{\tau}

\def\l{\lambda}

\def\e{\epsilon}

\def\bi{\begin{itemize}}
\def\ei{\end{itemize}}

\def\bp{\bar{\phi}}

\date{May 25th, 2008} \preprint{DFT-UAM-09-010\\IFT-UAM/CSIC-09-28}
\title{Eternity  and the   cosmological constant} \author{Enrique \'Alvarez, Roberto Vidal \\  Instituto de F\'{\i}sica Te\'orica
UAM/CSIC and Departamento de F\'{\i}sica Te\'orica \\ Universidad
Aut\'onoma de Madrid, E-28049--Madrid, Spain \\ E-mail: \email{enrique.alvarez@uam.es}, \email{jroberto.vidal@uam.es}}
\abstract{
The purpose of this paper is to analyze the stability of interacting matter in the presence of a cosmological constant. Using an approach based on the heat equation,  no imaginary part is found for the effective potential   in the presence of a fixed background, which is the n-dimensional sphere or else an analytical continuation thereof, which is explored in some detail.}

\begin{document}
\newpage
\section{Introduction}

A recurrent dream in theoretical physics is that a gravitational state with a nonvanishing cosmological constant is unstable. This idea was explicitly stated in \cite{Poly} more than a quarter century ago, but it is perhaps older. This has been predicated mainly in the context of de Sitter space, but if the stable gravitational state should be Minkowski space there is a clear need of a similar statement  
concerning negative values of the cosmological constant. Given the fact that there is some evidence that classically the constant curvature, maximally symmetric spaces are stable with respect to linear perturbations irrespectively of the sign of the curvature \cite{Abbott}, the purported instabilities must have quantum origin. The work of Abbott and Deser established positive Killing energy for small fluctuations of the gravitational field. In cases such as de Sitter space, in which there is no Killing vector which is globally timelike, the fluctuations have got to be contained {\em inside} the corresponding horizon. In Anti de Sitter space they were able to show stability with respect to all asymptotically vanishing fluctuations whatever large.
\par
The instability claim has been recently put on a new basis in a recent paper by Polyakov \cite{Polyakov} (where some references to earlier work can be found; many  that are not there can be found in the book \cite{Birrell}).
\par
It is well-known that there is a one-parameter family of so-called vacuum states in de Sitter space, first uncovered by Chernikov and Tagirov \cite{Chernikov}; a recent reference is \cite{Spradlin}. Ariadna's thread in this maze is usually taken as the strength and physical location of the singularities of the propagators (cf. \cite{Allen}). What is proposed in reference \cite{Polyakov} is to consider instead a different guiding principle, namely the "composition principle", a property which seems natural from the first quantized path integral approach to the theory. This property uniquely selects a particular propagator.
\par
This propagator is then used to claim that the presence of quantum fields interacting in de Sitter space an instability appears which manifest itself as an imaginary part of the free energy of the quantum fields. The stability is asociated to the concept of {\em eternity} \cite{Polyakov}. It is not fully clear to begin with, that this is the correct observable to consider, at least when there are horizons present (like in de Sitter space, in which no Killing energy can be globally defined). We shall nevertheless compute it, because it is anyway the first step towards more satisfactory calculations.
\par
 This claim is possibly related, but not identical, to the one put forward since quite a few years by Tsamis and Woodard \cite{Tsamis} and recently criticized by Garriga and Tanaka \cite{Garriga}. The latter is  a quantum gravitational effect; whereas the one we are considering in this paper is supposed to appear when considering quantum fields in a gravitational background, and  
neglecting backreaction effects.
\par
The relationship of these different claims with the status of de Sitter space as a vacuum of quantum gravity \cite{Witten} is not altogether clear. To the best of our knowledge, de Sitter space is at best a metastable solution of string theory \cite{Kachru}.
But the reason for that seems to rely strongly on gravitational interactions.
\par
The aim of the present paper is a quite modest one, namely to examine these assertions from a slightly different perspective, by studying the heat kernel which is nothing else than a particular solution of the heat equation, which is in turn a sort of euclidean version of Schr\"odinger's equation. This allows a straightforward determination of the free energy to one loop order. We follow the lead of the solution all the way down from the sphere towards its different analytical continuations.
The setup of the problem is then as follows. The free energy is given by a path integral over the gravitational fluctuations around a background $\bar{g}_{\m\n}$ as well as around fluctuations of the matter fields around their backgrounds $\bar{\phi}_a$, which are assumed to be solutions of the classical equations of motion. If  the gauge fixing is such that no mixing matter/gravity is generated, then the free energy is given to one loop order by a set of determinants.
\bea
&&W\left[\bar{g}_{\m\n},\bar{\phi}_a\right]\equiv \bar{S} 
\left(\bar{g}_{\m\n},\bar{\phi}_a\right)-\frac{1}{2}\textrm{tr}\,\log\, M^2_{\m\n 
\a\b}\left(\bar{g}_{\m\n},\bar{\phi}_a\right)+\nonumber\\
&&\textrm{tr}\,\log\,M_{gh}\left(\bar{g}_{\m\n},\bar{\phi}_a\right)-\frac{1}{2}\textrm{tr}\,\log\, M^2\left(\bar{g}_{\m\n},\bar{\phi}_a\right)
\eea
where $ M^2_{\m\n\a\b}\left(\bar{g}_{\m\n},\bar{\phi}_a\right)$ represents the quadratic operator acting on gravitational fluctuations using a background gauge fixing, $M_{gh}\left(\bar{g}_{\m\n},\bar{\phi}_a\right)$ the corresponding operator for the ghosts and finally, $ M^2\left(\bar{g}_{\m\n},\bar{\phi}_a\right)$ stands for the quadratic operator for the matter fields.
\par
Assuming, for simplicity, that all matter is composed by scalar fields, and neglecting the dynamics of the gravitational field, id est 
\bea
&&S_m\equiv \int d^n x \sqrt{|\bar{g}|}\frac{1}{2}\bar{g}^{\m\n}\d_{ab} 
\pd_\m\phi^a\pd_\n\phi^b-\sum_a\xi \bar{R}\left(\phi^a\right)^2-V(\phi^a)=\nonumber\\
&&\bar{S}_m-\frac{1}{2}\int d^n x \sqrt{|\bar{g}|}\phi^a\,\left. \pd_a 
\pd_b V\right|_{\bar{\phi}}\, \phi^b + O\left(\phi^2\right)
\eea
that is, the operator that interests us is, in an obvious notation,
\be
M^2\left(\bar{g}_{\m\n},\bar{\phi}_a\right)\equiv -\bar{\Delta}\d_{ab}- 
\bar{ V}_{ab}
\ee
Generically, we are only able to compute it in the approximation where the background scalar field is constant; that is, we are evaluating the {\em effective potential}.

\section{The Composition law}
It is well known (cf. for example the discussion in \cite{Polya}) that in flat space the Klein-Gordon propagator can be recovered from the first quantized path integral
\[G(x,y)\equiv \int {\cal D} X(s)e^{- m S(X)}\]
where the integral extends to all paths such that
\bea
&&X(0)=x\nonumber\\
&&X(1)=y
\eea
and the action for each path is
\[S(X)\equiv \int_0^1 d\t \sqrt{\delta_{\m\n}\dot{X}^\m\dot{X}^\n}\]

This representation makes manifest that the propagator enjoys a quantum mechanical composition law, at least in the euclidean case:
\be
\int d^n z\,G(x,z)G(z,y)=\int d^n z\,{\cal D}X(s){\cal D}Y(s)  e^{-m\{S(X)+S(Y)\}}
\ee
where $X(s)$ goes from $x$ to $z$ and $Y(s)$ from $z$ to $y$.
Then
\be
\int d^n z G(x,z)G(z,y)=\int  {\cal D}X(s) e^{-m S(X)} {\cal F}\left(m^2,S(X)\right)
\ee
where now $X(s)$ goes from $x$ to $y$, and the extra factor ${\cal F}\left(m^2,S(X)\right)$ takes into account the integral over the intermediate point $z$ along the curve and leads to
\begin{equation}\label{composition}
\int d^n z\,G(x,z)G(z,y)=-\frac\partial{\partial m^2}G(x,y)
\end{equation}
(This is equivalent to assert that ${\cal F}\left(m^2,S(X)\right)= \frac{1}{2 m} S(X)$. We are aware of no simple argument for this).

In a recent paper Polyakov \cite{Polyakov} suggests that unitarity in quantum field theory is equivalent to this {\em path composition}. Asymptotically (for large separation between the points) the propagator should behave  as
\be
G(x,y)\sim e^{-i m s(x,y)}
\ee
where $s(x,y)$ is the geodesic distance between the points $x$ and $y$.
\par
The flat space Klein-Gordon propagator can be easily recovered \cite{Polya} through
\footnote{In flat space this identity is true in any dimension for true propagators (id est, solutions of the inhomogeneous equation)
because using the Fourier representation
\be
G(x,y)=\int\frac{d^n p}{(2\pi)^n} \frac{e^{ip(x-y)}}{p^2+m^2}
\ee
and 
\be
\int d^n z G(x,z)G(z,y)=\int d^n z \frac{d^n p}{(2\pi)^n}\frac{d^n k}{(2\pi)^n}\frac{e^{ip(x-z)}}{p^2+m^2}\frac{e^{ik(z-y)}}{k^2+m^2}=-\frac{\pd}{\pd m^2}G(x,y)
\ee
Direct verification is more laborious.}
\begin{equation}\label{GcomoK}
G(x,y)=\int_0^\infty d\t K(\t;x,y)
\end{equation}
where $K(\t;x,y)$ is the Schr\"odinger functional
\begin{equation}
K(\t;x,y)\equiv \int{\cal D}X e^{-i\int^\t_0 d\sigma\left(\frac{\dot{X}^2}{2\sigma}+\sigma \frac{m^2}{2}\right)}
\end{equation}
and $\t$ is the gauge invariant distance $\t\equiv \int_0^1 e(\lambda)d\lambda$. Polyakov's path composition is then a simple consequence of Feynman's kernel quantum mechanical composition law
\begin{equation}
\int d^n z K\left(\t_1;y,z\right)K\left(\t_2;z,x\right)=K\left(\t_1+\t_2;y,x\right)
\end{equation}
Once these facts are understood, the temptation to choose them as the starting point for the study of quantum fields in a gravitational background is irresistible.\\

The preceding results are by no means restricted to flat space. We shall explain in a moment that
given the heat kernel, that is, the  solution of  the heat equation in an arbitrary spacetime $\displaystyle \partial_\tau K=(\Delta-m^2) K$ with the initial conditions $K(0;x)=\delta(x)$ we can obtain a Green's function for the Klein-Gordon equation through
\begin{equation}
G(x)=\int_0^\infty \,K(\tau;x)\,d \tau=\int\theta(\tau)\,K(\tau;x)\,d \tau
\end{equation}
\[
(\Delta-m^2) G(x)=\int_0^\infty \,(\Delta-m^2) K(\tau;x)\,d\tau=
\]
\begin{equation}
=\int_0^\infty \,\partial_\tau K(\tau;x)\,d\tau=K(\tau;x)\Big|^\infty_0=-\delta(x)
\end{equation}

Whenever the composition principle of Schr\"odinger (or the heat) equation holds
\begin{equation}
\int K(\tau;x,z)K(\sigma;z,y)\,d^nz=K(\tau+\sigma;x,y)
\end{equation}
this propagator (and others related) enjoys automatically the composition law (\ref{composition})
\[
\int G(x,z)G(z,y)\,d^nz=\int_C dtds\,K(t;x,z)K(s;z,y)\,d^nz=
\]
\begin{equation}
=\int_C dt dsK(t+s;x,y)=\frac12\int_{C'} d\tau d\sigma\,K(\tau;x,y)\textrm{ ,}
\end{equation}
where the integration domain in the $t,s$ plane is the upper right quadrant $C$. We have performed the transformation $\tau=t+s$, $\sigma=t-s$, and the new domain $C'$ can be parametrized as
\begin{equation}
\frac12\int d\tau d\sigma\,\theta(\tau+\sigma)\theta(\tau-\sigma)\,K(\tau;x,y)=\int d\tau\,\tau \theta(\tau)K(\tau;x,y)=-\partial_{m^2}G(x,y)
\end{equation}
where we take in account that the heat kernel for mass $m$ is related to the massless one by $K_{m^2}=e^{-m^2 \tau}K_{m=0}$. The conclusion of the above is that starting from the heat kernel, the ``composition principle" is a simple consequence of the
quantum mechanical closure relation
\be
\sum_z |z\rangle\langle z|=1
\ee
\section{The heat kernel}
What we shall denote by {\em heat kernel} is what mathematicians call the {\em fundamental solution of the real heat equation} (FSRHE) made popular by Kac when he asked the question as to whether one could hear the shape of a drum \cite{Kac}
(the short answer is that one cannot in general). The mathematicians call heat equation to
\[\Delta K(x,y;\t)-\m^2\frac{\pd K(x,y;\t)}{\pd \t}=0\]
where $\Delta\equiv \nabla_\m\nabla^\m$, and we have introduced a mass scale $\m$ to make $\t$ dimensionless (or, what is equivalent, to consider the operator $\frac{\Delta}{\m^2}$, whose eigenvalues
are also dimensionless).
The FSRHE is defined as the solution such that $\displaystyle\lim_{\t\to 0^+}K(x,y;\t)=\d(x,y)$. The importance of the FSRHE is that it is {\em unique} for compact connected $C^\infty$ riemannian manifolds without boundary
\cite{Berger}. Formally, it can be predicated that
\[K(\t)\equiv e^{\frac{\t}{\m^2} \Delta}\]
(the convention is that the operator in the exponent is negative definite for $\t\in \mathbb{R}^+$.)
so that a Green's function can be defined as
\[G\equiv -\Delta^{-1}\equiv \int_0^\infty K(\t)d\t\]
This Green's function is also unique under the same conditions than the FSRHE is.
\par
We will deal with this equation with an additional mass term, as in the previous section. In the particular case of euclidean space $\mathbb{R}^n$ (which is non compact, by the way)
\[K_0\left(x,y;\t\right)=\frac{\m^{n-2}}{\left(4\pi\t\right)^{n/2}}e^{-\frac{\m^2\left(x-y\right)^2}{4\t}-\frac{m^2}{\m^2}\t}\]
(where $\m$ is an arbitrary mass scale whose physical meaning is the same as the one appearing in dimensional regularization).
The famous integral
\be
\int_0^\infty dx\, x^{\n-1}e^{-\frac{\b}{x}-\g x}=\left(\frac{\b}{\g}\right)^{\n/2}\,K_\n\left(2\sqrt{\b\g}\right)
\ee
leads to the euclidean Green's function
\[G_0\left(x,y\right)\equiv \int_0^\infty  d\t K_0\left(x,y;\t\right) =\frac{1}{2\pi}\left(\frac{m}{2\pi|x-y|}\right)^{n/2 -1}K_{n/2 -1}\left(m|x-y|\right)\]
where $|x|^2\equiv \sum_1^n x_i^2$ and $K_n(x)$ is the Bessel function of imaginary argument. This is the mother of all Green's functions.\\

This whole procedure can in some sense be reversed. If we consider the heat kernel corresponding to the {\em massless} Klein-Gordon operator, $K_{m=0}(\t)\equiv K(\t)e^{\frac{m^2}{\m^2} \t}$, then the relationship between the heat kernel and the (massive) Green's function is just a Laplace transform
\[G_{m}(x)=\int_0^\infty K_{m=0}(\t) e^{-\t \frac{m^2}{\m^2}}d\t\]

This means that whenever the Green's function as a function of $m^2$ is bounded by a polynomial in the half plane $\textrm{Re}\,m^2 \geq c$, the Laplace transform can be inverted to yield
\[K_{m=0}(\t)=\frac{1}{\m^2}\int_{c-i\infty}^{c+i\infty} d m^2 e^{\t \frac{m^2}{\m^2}} G_{m}(x)\]\\
\par

We shall extend this precise and beautiful mathematical framework in two ways. First of all, physics forces upon us the consideration of operators somewhat more general than the covariant  laplacian, for example by allowing a generalized mass term (as well as nonminimal operators for higher spins \cite{Barvinsky}). Secondly, we are eventually interested in {\em pseudo}-riemannian, Lorentzian geometries which are moreover non-compact.\\

One of our main worries in the present paper will precisely be how to go back and forth from one signature to the other. What we have seen in the previous paragraph is that this particular Green's function also satisfies Polyakov's composition principle.\\

The class of spaces we are going to be interested at in this paper are all related to the sphere by analytic continuation.
The sphere $S_n$ itself can be defined as the compact form of the symmetric space $SO(n+1)/SO(n)$. It can also be usefully defined
as the hypersurface 
\be\label{definingsphere}
\sum_{A=0}^n  X_A^2\equiv \d_{AB}X^A X^B = l^2
\ee
on a flat $\mathbb{R}_{n+1}$ space\footnote{These coordinates, which we are going to represent in capital letters, are usually denoted as {\em Weierstrass} coordinates.} with metric $ds^2= \d_{AB}dX^A dX^B$; or else the real projective space, $\mathbb{R P}_n=S_n/\mathbb{Z}_2$, where $\mathbb{Z}_2$ is the antipodal mapping
\be
\mathbb{Z}_2:X^A\rightarrow -X^A
\ee
The sphere is then the universal covering space of the projective plane, and $\pi_1(\mathbb{RP}_n)=\mathbb{Z}_2$. Functions on the projective plane are given by even functions on the sphere
\be
f(X^A)=f(-X^A)
\ee
The projective plane is non-orientable for even values of n, but it is orientable for odd values of n. For example, $\mathbb{RP}_1\sim S_1$.\\

\vspace{1cm}

In their work on the Schr\"odinger equation, Grosche and Steiner \cite{Grosche} are led towards the following integral, which gives what is essentially the Schr\"odinger propagator:
\bea
&&K\left(\Omega,\Omega^\prime;\tau\right)\equiv \int {\cal D}\Omega\,e^{i\int_0^\tau d\lambda\left(\frac{m l^2}{2}\dot\Omega^2+\frac{n(n-2)}{8 m l^2}\right)}=e^{i\tau\frac{n(n-2)}{8 m l^2}}\int {\cal D}\Omega\,e^{i\int_0^\tau d\lambda\frac{m l^2}{2}\dot\Omega^2}\equiv \nonumber\\
&&e^{i\tau\frac{n(n-2)}{8 m l^2}}Z\left(\Omega,\Omega^\prime;\tau\right)
\eea
where $\Omega\equiv\vec{n}$ is a unit vector, defining a point on the unit sphere $\vec{n}\in S_{n}$, and can be characterized in polar coordinates by a set of angles, $\theta_1\ldots \theta_n$.
\par
The path integral will be done by means of Feynman's time slicing technique.
The action reads
\be
S=\frac{m l^2}{2}\sum_{i=1}^n\left(\vec{\Omega}_i-\vec{\Omega}_{i-1}\right)^2=m l^2\sum_{i=1}^n\left(1-\cos\,\psi_{i-1}\right)
\ee
where we have defined
\be
\cos\,\psi_{i-1}\equiv\vec{\Omega}_i\cdot\vec\Omega_{i-1}
\ee
The expansion discussed in the appendix conveys the fact that
\be
e^{z\,\cos\,\psi}=\left(\frac{z}{2}\right)^{-\frac{n-1}{2}}\Gamma\left(\frac{n-1}{2}\right)\sum_{j=0}^\infty\left(j+\frac{n-1}{2}\right)I_{j+\frac{n-1}{2}}\left(z\right)C_j^{\frac{n-1}{2}}\left(\cos\,\psi\right)
\ee

\begin{equation}
Z\left(\theta,\theta^\prime;\tau\right)=e^{i\tau\frac{n(n-2)}{8 m l^2}}\int {\cal D}\Omega\,e^{i\int\frac{m l^2}2\dot\Omega^2}=e^{i\tau\frac{n(n-2)}{8 m l^2}}\int \prod_i d\Omega_i\,e^{im l^2\sum_i\left(1-\cos\,\psi_{i-1}\right)}
\end{equation}
the integrations to be done are, schematically,
\bea
&&\int d\Omega_1\ldots d\Omega_{n-1} \sum_{j_1\vec m_1}\sum_{j_2\vec m_2}\ldots Y_{j_1\vec{m_1}}(\Omega_1) Y^*_{j_1\vec{m_1}}(\Omega_0) Y_{j_2\vec{m_2}}(\Omega_2)Y^*_{j_2\vec{m_2}}(\Omega_1)\ldots\nonumber\\
&&\ldots\sum Y_{j_n\vec{m_n}}(\Omega_n)Y^*_{j_n\vec{m_n}}(\Omega_{n-1})=\sum_{j\vec m} Y_{j\vec{m}}(\Omega_n)Y^*_{j\vec{m}}(\Omega_{0})\sim\sum_j C^{\frac{n-1}{2}}_j(\cos\,\psi)\nonumber\
\eea

The final result of \cite{Grosche} is

\be
K\left(\Omega,\Omega^\prime;\tau\right)=\frac1{V(S_n)}\sum_{j=0}^\infty \frac{2j+n-1}{n-1}C_j^{\frac{n-1}{2}}(\Omega\cdot\Omega^\prime)e^{-\frac{i \tau}{2 m l^2}j(j+n-1)}
\ee

Our main tool in order to study the effective potential in constant curvature spaces will be the analogous of the preceding computation for our Klein-Gordon equation, as well as the representation of the delta function on the sphere $S_{n-1}$ by means of Gegenbauer polynomials (cf. Appendix) , id est,

\begin{equation}\label{heatkernel}
K(\tau;\Omega,\Omega')=\frac 1{V(S_n)}\sum_j\frac{n-1+2j}{n-1}\,C^\frac{n-1}{2}_j(\Omega\cdot\Omega')e^{-\tau(m^2l^2+j(j+n-1))}
\end{equation}

that is the solution of the heat equation such that

\be
\lim_{\t\rightarrow 0^+}K(\t;\Omega,\Omega^\prime)=\d\left(\Omega-\Omega^\prime\right)
\ee
where the delta function reads
\begin{equation}
\delta(\Omega-\Omega')=\frac 1{V(S_n)}\sum_j\frac{n-1+2j}{n-1}\,C^\frac{n-1}{2}_j(\cos\theta_n)
\end{equation}

\vspace{1cm}

We can see the heat kernel formally as
\be
K(\t)\equiv e^{-\t \bar{M}^2}
\ee
where $\bar{M}^2$ is the positive definite operator acting on quadratic fluctuations around the background field, id est,
\be
\bar{M}^2\equiv -\Delta+ \pd^2 V(\bp)
\ee
and we include masses in the potential.

Let us mention that whenever the full eigenvalue problem for the operator $\bar{M}^2$ is known, there is a formal FSRHE. Using the discrete notation,
\be
\bar{M}^2 u_n(x)=\l_n u_n(x)
\ee
with eigenfunctions which can be chosen to obey
\be
\left(u_n,u_m\right)\equiv \int d \m(x)\, u^*_n(x) u_m(x)=\d_{nm}
\ee
(where the measure $d\m(x)$ is usually $\sqrt{|g|} d^n x $) as well as  a completeness relationship of the type
\be
\sum_n u^*_n(x) u_n(y)=\d(x-y)
\ee
then the following is the sought for FSRHE 
\be
K(x,y|\t)=\sum_n e^{-\l_n \t}u^*_n(x) u_n(y)
\ee
whose imaginary part is determined by the one of the eigenvalues themselves.\\
\par

As we have already advertised, in order to study the free energy up to one loop order, it is much more convenient to study the heat kernel,
than the Green's function, because it gives the desired result directly

\be\label{freeenergy}
W=\frac{1}{2}\int_0^\infty \frac{d\t}{\t}\textrm{tr}\,\int d^n x \sqrt{|g|} K\left(\t;x,x\right)
\ee
This definition  {\em includes}  the definition based to the zeta-function (which is the finite part) as well as the divergent counterterms.
\par

Before that, however, let us clarify a few points on the relationship between Green's functions in constant curvature spaces.
Although the defining equations of the different spaces themselves in Weierstrass coordinates are analytic continuations of the equation of the sphere,  some subtleties  appear with the analytic continuation of Green's functions.

\section{Green's functions in constant curvature spaces.}

We shall mainly be concerned in this paper with fundamental solutions of the Klein-Gordon equation in the real sections of the sphere, invariant under the full group of isometries. Related analysis have been performed in \cite{Camporesi}\cite{AllenJacobson}. The homogeneous version of this equation takes always the same form in these spaces:
\begin{equation}
(z^2-1)G''+nzG'\pm m^2l^2=0
\end{equation}
where $z$ is the corresponding geodesic distance for each space (cf. \ref{zeta}).\\

The problem of finding the invariant Green's functions of this equation can be solved in a simple and general way. The full space of solutions is two-dimensional. All we have to do is extending the domain of definition of these functions to the appropiate region of the real axis for each surface.\\

We have to take care also of the singularities we obtain. We are interested in a single source (tipically in the ``north pole'' $z=1$), or perhaps in symmetric solutions under $\mathbb Z_2$ in order to obtain Green's functions for the projective case.\\

\begin{center}
\begin{figure}[h]\label{ruta}
\includegraphics[scale=0.6]{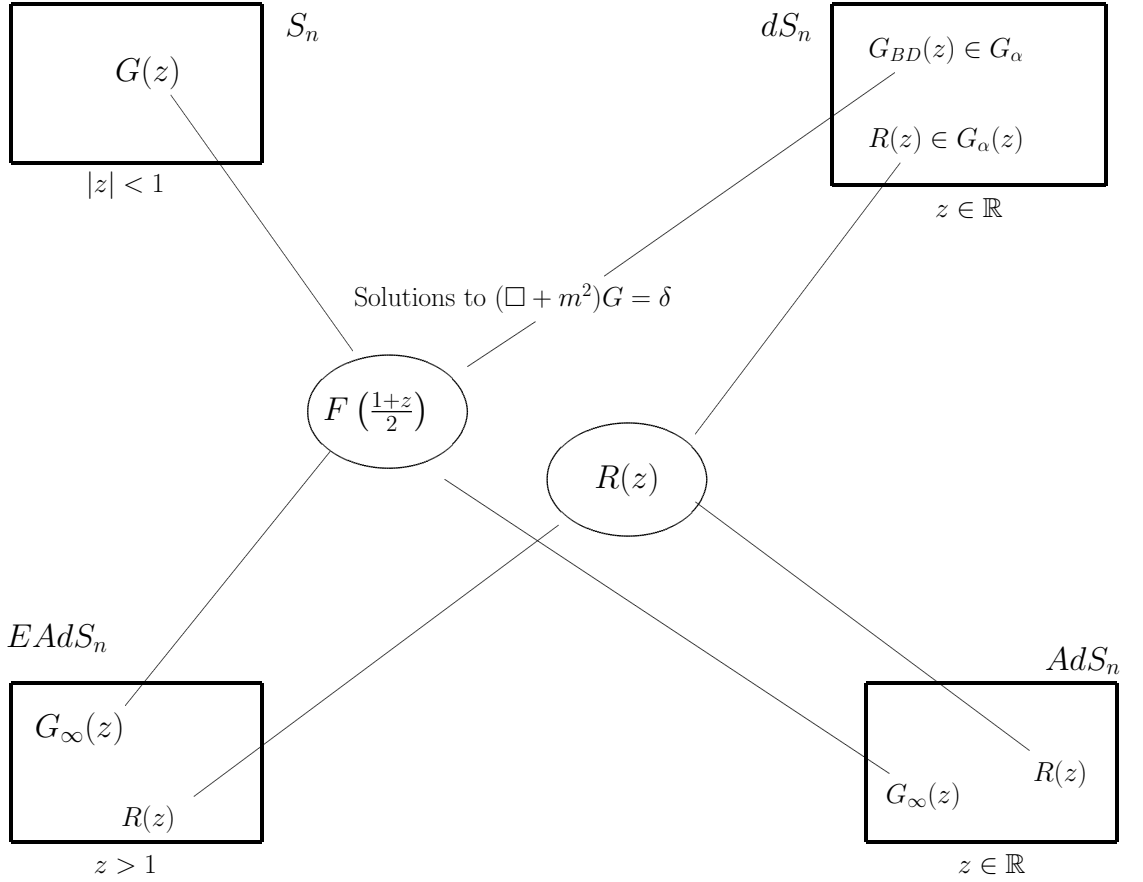}
\caption{Route sheet of analytic continuations.}
\end{figure}
\end{center}

In the Fig. \ref{ruta} we have summarized the results. Combining solutions of the generic Klein-Gordon equation (hypergeometric functions) with the appropriate singularity ($F\left(\frac{1+z}2\right)$, $R$), we can build several different propagators for each space. Here $R$ is proportional to a Legendre $Q$ function, finite at $z=\infty$. $G_\infty$ means a Green's function that diverges at infinity. $G_\a$ stands for the Green's functions of the $\a$-vacua.

\subsection{Flat spacetime}
The flat spacetime case is interesting in order to know the appropriate short distance behaviour. We saw in the previous that the calculation of the n-dimensional Green's function in an euclidean flat spacetime gives 
\begin{equation}
G(x)=\int\frac{e^{ipx}}{p^2+m^2}\,\frac{d^np}{(2\pi)^n}=\frac1{2\pi}\left(\frac m{2\pi r}\right)^{\frac n2-1}K_{\frac n2-1}(mr)
\end{equation}

When we perform the analytic continuation to the Feynman propagator in lorentzian signature, we implicitly chose the prescription such that the result is still a propagator, i.e. that keeps the appropriate singularity:
\[
G_F(x)=\frac i{2\pi}\left(\frac m{2\pi\sqrt{-x^2+i\epsilon}}\right)^{\frac n2-1}K_{\frac n2-1}(m\sqrt{-x^2+i\epsilon})
\]
That this is correct, can be checked performing the integral $\displaystyle\int\frac{d^nk}{(2\pi)^n}\frac{e^{ikx}}{-k^2+m^2-i\epsilon}$ explicitly. The branch cut of $\sqrt{-x^2}$ does not depend on the sign on time, but just on $|t|$, as was expected from a time ordering.\\

The singularity of this propagator is:
\begin{equation}\label{coef}
G(x)\xrightarrow{x^2\to0}\frac i{(2\pi)^\frac{n}{2}}2^{\frac n2-2}\Gamma\left(\frac n2-1\right)(-x^2+i\epsilon)^{1-\frac n2}+[\log(-x^2+i\epsilon)]
\end{equation}
where the term in brackets appears when $n$ is even.\\

This prescription precisely gives us the correct singularity to recover a delta function. Other possibilities lead to homogeneous solutions which correspond to important functions:
\begin{itemize}
\item Wightman function $-i	W$: $x^2\to -x^2+i\epsilon t$
\item Symmetric function $G^{(1)}$: Re $W$
\item Pauli-Jordan function (conmmutator) $D$: Im $W$
\end{itemize}

\subsection{Sphere}
In the appendix we give some details on different metrics for constant curvature spaces with different signatures.
The Klein-Gordon equation in the n-dimensional sphere reads:
\begin{equation}
\frac1{\sin\theta^{n-1}}\partial_\theta(\sin\theta^{n-1}\partial_\theta G)-m^2l^2G=0=\frac1{(1-z^2)^{\frac{n-2}2}}\partial_z((1-z^2)^{\frac n2}\partial_zG)-m^2l^2G
\end{equation}
where $z=\cos\theta$. This is almost an hypergeometric equation:
\begin{equation}
(z^2-1)G''+nzG'+m^2l^2G=0
\end{equation}
with the solutions\footnote{The possible values of $\mu$ are real and positive, or imaginary, with $\frac{n-1}2>-i\mu>0$}:
\begin{equation}\label{soluciones}
G(z)=F_\pm(z)=F\left(\frac{1\pm z}2\right)\equiv F\left(i\mu+\frac{n-1}2,-i\mu+\frac{n-1}2;\frac n2;\frac{1\pm z}2\right)
\end{equation}
where $m^2l^2=\mu^2+\frac{(n-1)^2}4$. Each one is singular respectively in $z=\pm1$, and this singularity corresponds precisely to delta function in opposite points. in this way we recover the well known fact that there is a single Green's function in the sphere.\\

The composition law holds for this Green's function, given that is unique and therefore, proportional to the alternate expression:
\begin{equation}
G(\Omega\cdot\Omega')=\sum_{j\vec k}\frac{Y_{j\vec k}(\Omega)Y_{j\vec k}(\Omega')^*}{j(j+n-1)+m^2}
\end{equation}
given in terms of eigenfunctions of $\Delta$, i.e. spherical harmonics, and their eigenvalues. It is straightforward to check the composition law with this formula.

\subsection{de Sitter space} 
The Klein-Gordon equation in this case reads
\[
\frac 1{\cosh\tau^{n-1}}\partial_\tau\left(\cosh\tau^{n-1}\partial_\tau G\right)-\frac1{\cosh\tau^2\sin\theta^{n-2}}\partial_\theta\left(\sin\theta^{n-2}\partial_\theta G\right)+m^2l^2G=0
\]
\begin{equation}
(z^2-1)G''+nzG'+m^2l^2G=0\textrm{ , }z=\cosh\tau\cos\theta 
\end{equation}

The solution is given by the same expression as before. In order to provide a function defined over the full de Sitter space (for all $z\in\mathbb R$), we must specify the values in the branch cuts. In addition, since the signature of spacetime has changed, this prescription will determine the character of the singularity, i.e. homogeneous or not.\\

Looking to the flat spacetime case, the solution is simple, since the short distance behaviour should match. The correct analytic continuation is:
\begin{equation}\label{realz1}
G_{BD}(z)=F\left(i\mu+\frac{n-1}2,-i\mu+\frac{n-1}2;\frac n2;\frac{1+z}2-i\epsilon\right)
\end{equation}
and this is (proportional to) the euclidean or Bunch-Davies propagator. In addition we can continue the both solutions in such a way that they remain homogeneous, for example:
\begin{equation}\label{realz2}
\textrm{Re}F_\pm(z)=\textrm{Re}\,F\left(i\mu+\frac{n-1}2,-i\mu+\frac{n-1}2;\frac n2;\frac{1\pm z}2\right)
\end{equation}
where we denote by $\textrm{Re},i\textrm{Im} f(z)=f(z+i\epsilon)\pm \,f(z-i\epsilon)$. This combination cancels the delta divergence.\\

The above expression spans the space of homogeneous invariant solutions that originates the ambiguity in the propagator:
\begin{equation}\label{general}
G(z)=G_{BD}(z)+\alpha\,\textrm{Re}\,F_+(z)+\beta\,\textrm{Re}\,F_-(z)
\end{equation}
However, if the propagator comes from a vacuum expectation value, we know \cite{Allen} that just a 1-parameter family survives, the $\alpha$ ($\alpha>0$) vacuum\footnote{The most general expression, de Sitter invariant except for the discrete symmetries, is the $\alpha,\beta$ vacuum, with $\beta\in[0,2\pi)$:
\begin{equation}
G_{\alpha,\beta}(x,y)=\frac{i|\Gamma\left(i\mu+\frac{n-1}2\right)|^2}{2(4\pi)^\frac{n}{2}\{-\Gamma(2-\frac n2)|\Gamma(\frac n2)\}}\Bigg\{\cosh2\alpha\,\textrm{Re}F\left(\frac{1+z}2\right)+\qquad\qquad\nonumber
\end{equation}
\begin{equation}
+\sinh2\alpha\left[\cos\beta\,\textrm{Re}F\left(\frac{1-z}2\right)-\sin\beta\,\sigma\textrm{Im}F\left(\frac{1-z}2\right)\right]-i\,\textrm{Im}F\left(\frac{1+z}2\right)\Bigg\}\nonumber
\end{equation}
where $\sigma$ is the sign of the time-ordering of $(x^A,y)$. This is defined only in the case $z<-1$, but for $z>-1$ the imaginary part of $F\left(\frac{1-z}2\right)$ vanishes, as in the case of the conmmutator function. This expression for $\beta\neq0$ is not fully de Sitter invariant, i.e. it does not depend only on $z$, due precisely to the presence of this sign.
}:
\begin{equation}
G_\alpha(z)=\frac{i|\Gamma\left(i\mu+\frac{n-1}2\right)|^2}{2(4\pi)^\frac{n}{2}\{-\Gamma(2-\frac n2)|\Gamma(\frac n2)\}}\Bigg\{\cosh2\alpha\,\textrm{Re}F\left(\frac{1+z}2\right)+\qquad\qquad\nonumber
\end{equation}
\begin{equation}
\qquad\qquad+\sinh2\alpha\,\textrm{Re}F\left(\frac{1-z}2\right)-i\,\textrm{Im}F\left(\frac{1+z}2\right)\Bigg\}
\end{equation}

The term in the $\{|\}$ corresponds to the \{odd$|$even\} case.\\

\subsection{Euclidean Anti de Sitter space}
Now the Klein-Gordon equation reads
\begin{equation}\label{KGm}
(z^2-1)G''+nzG'-m^2l^2G=0
\end{equation}
The solutions are pretty similar to the sphere case:
\begin{equation}
G(z)=F\left(\mu+\frac{n-1}2,-\mu+\frac{n-1}2;\frac n2;\frac{1\pm z}2\right)
\end{equation}
where $\mu^2=m^2l^2+\left(\frac{n-1}2\right)^2$. This time $\mu>\frac{n-1}2$.\\

The negative sign solution is regular in $z=1$ so it is purely homogeneous. Given that now $z\geq 1$, the positive sign solution needs a prescription in the branch cut to be meaningful. The exact behaviour near $z=1$ depends on the parity of $n$, but in both cases the expressions are like:
\begin{equation}
F\left(\frac{1+z}2\right)=\ldots+\ldots\cdot\left(\frac{1-z}2\right)^{1-\frac n2}
\end{equation}
where $\ldots$ something regular in $z=1$ (or a logarithm). We can see from this equation that taking the upper or lower limit in the real axis, $z\pm i\epsilon$ gives us a Green's function $G_\infty$.\\

However, this propagator $G_\infty$ diverges in the infinity, as we can see from the expansion of the hypergeometric function near the infinity:
\begin{equation}
F(\alpha,\beta;\gamma;z)\xrightarrow{z\to\infty}\textrm{const}\,(-z)^{-\alpha}+\textrm{const}\,(-z)^{-\beta}
\end{equation}
from wich we get:
\begin{equation}
G_\infty(z)\xrightarrow{z\to\infty}\textrm{const}\,\left(-\frac{1+z}2\right)^{-\mu-\frac{n-1}2}+\textrm{const}\,\left(-\frac{1+z}2\right)^{\mu-\frac{n-1}2}
\end{equation}
Both the imaginary and the real part of this expression diverge (this is due to the second term), so in general no prescription gives us a propagator that vanishes at infinity\footnote{In fact, some specific values of $m$ are such that taking only the imaginary [real] part of the function, for $n$ odd [even], this term is cancelled.}.\\

An appropiate solution can be obtained combining the $G_\infty$ with the homogeneous solutions. The exact expression can be given in terms of Legendre associated functions:
\begin{equation}\label{R}
G(z)=(z^2-1)^{\frac{1-n}4}\,Q^\frac{n-1}{2}_{\mu-\frac12}(z)\sim z^{-\mu-\frac{n-1}2}F\left(\frac\mu2+\frac{n+1}4,\frac\mu2+\frac{n-1}4;\mu+1;\frac1{z^2}\right)
\end{equation}
This special combination, that we will abbreviate $R^\frac{n-1}{2}_{\mu-\frac12}$, is a solution of (\ref{KGm}). The composition principle holds for this propagator, given that this solution is the Laplace transform of the Schr\"odinger propagator of $EAdS$ \cite{Grosche}. 

\subsection{Anti de Sitter space} The Klein-Gordon equation in $AdS$ is identical to the $EAdS$ case. The variable $z$ can take any real value again, as in de Sitter, so the the solutions to (\ref{KGm}) can be continued in the same way as in (\ref{realz1}), (\ref{realz2}). We have just to take in account that now $i\mu\to\mu$, where $\mu$ means the same as in the $EAdS$ case.\\

Since the Anti de Sitter space has a well defined spatial infinity at $z=\infty$, if we require the propagator to vanish there, we will obtain the same $R$ expression as in the $EAdS$ case (\ref{R}). However, in this case we have to extend the domain to the full real axis. In order to get the correct prescription, we need the relationship between the $R$ and the hypergeometric solutions:\\

\begin{equation}
R_\nu^{\frac{n-2}2}(z)=\rho_{n,\nu}\left\{e^{\mp i\pi\nu}F\left(\frac{1-z}2\right)+\varphi_\pm\,F\left(\frac{1+z}2\right)\right\}
\end{equation}
\[
\rho_{n,\nu}=\frac{2^{-\frac n2}\pi\Gamma(\frac n2+\nu)}{\Gamma(\frac n2)\Gamma(2-\frac n2+\nu)\{i\cos\pi\nu|\sin\pi\nu\}}\textrm{ ; }\varphi_\pm=\{i(-1)^{\frac{n\pm1}2}|(-1)^\frac{n}{2}\}	
\]
where again we write togheter the \{odd$|$even\} case, and the upper (lower) sign is for positive (negative) imaginary part of $z$.\\

An expression like (\ref{general}) is the most general Green's function. Since the delta singularities are in the imaginary part of the $F$ solutions, and the homogeneous pieces are the real parts, we have to eliminate the imaginary part of $F_-\equiv F\left(\frac{1-z}2\right)$, and it is easy to see that the appropriate combination to achieve it is 
\begin{equation}
\tilde R_\nu^\frac{n-2}{2}(z)=e^{i\pi\nu}R_\nu^\frac{n-2}{2}(z+i\epsilon)+e^{-i\pi\nu}R_\nu^\frac{n-2}{2}(z-i\epsilon)
\end{equation}
The detailed expressions in the even and odd cases are respectively:
\begin{equation}
\tilde R_\nu^\frac{n-2}{2}(z)\sim\textrm{Re}F_-(z)+(-1)^\frac{n}{2}\cos\pi\nu\,\textrm{Re}F_+(z)-(-1)^\frac{n}{2}\sin\pi\nu\,\textrm{Im}F_+(z)=\nonumber
\end{equation}
\begin{equation}
=\textrm{Re}F_-(z)+(-1)^\frac{n}{2}i\sinh\pi\mu\,\textrm{Re}F_+(z)+(-1)^\frac{n}{2}\cosh\pi\mu\,\textrm{Im}F_+(z)
\end{equation}
\[
\tilde R_\nu^\frac{n-2}{2}(z)\sim\textrm{Re}F_-(z)+(-1)^\frac{n-1}{2}\sin\pi\nu\,\textrm{Re}F_+(z)+(-1)^\frac{n-1}{2}\cos\pi\nu\,\textrm{Im}F_+(z)=
\]
\begin{equation}
=\textrm{Re}F_-(z)-(-1)^\frac{n-1}{2}\cosh\pi\mu\,\textrm{Re}F_+(z)+(-1)^\frac{n-1}{2}i\sinh\pi\mu\,\textrm{Im}F_+(z)
\end{equation}

The second line in each case come from $\nu=i\mu-\frac12$, i.e. the de Sitter case. As we can see, if and only if the dimension $n$ is odd the $R$ solution can be analitically continued into an alpha-beta vacuum, because of the inappropiate $i$ factors in the even case. The parameters of that vacuum are $\sinh2\alpha=\textrm{csch}\pi\mu$, and $\beta=0$ ($\beta=\pi$) for $(-1)^\frac{n+1}{2}$ positive (negative)\footnote{This is valid only in the case of $m>\frac{n-1}2$ in de Sitter. For lower masses there is no possibility of analytic continuation, because of the $i$ factors again.}. 

\subsection{Projective spaces}
A function defined over the projective version of these spaces can always be lifted to an symmetric function defined over the original space. It is very easy to obtain the most general Green's function of such an space, given the previous classification.\\

For the projective plane $\mathbb RP_n=S_n/\mathbb Z_2$, there is a single Green function corresponding to the projection of $G(z)+G(-z)$, where $G(z)$ is the propagator in \ref{soluciones} with the positive sign.\\

In the projective versions of de Sitter or Anti de Sitter, $dS_n/\mathbb Z_2$ and $AdS_n/\mathbb Z_2$, we found that the most general Green's function is:
\begin{equation}
G(z)=G_{BD}(z)+\alpha\,\textrm{Re}\,F_+(z)+\beta\,\textrm{Re}\,F_-(z)
\end{equation}
where $\alpha$ and $\beta$ are arbitrary constants. If we symmetrize this expression, we get the general propagator for these spacetimes:
\begin{equation}
G_P(z)=G_{BD}(z)+G_{BD}(-z)+\alpha\,(\textrm{Re}\,F_+(z)+\textrm{Re}\,F_-(z))
\end{equation}
In particular, we can symmetrize the $\tilde R$ solution finite at $z=\pm\infty$.

\section{The imaginary part of the effective potential.}
In flat space there is a systematic way of determining the ground state of a physical system, namely, to minimize the effective potential (the effective action for constant backgrounds). This is the physical principle that generalizes minimization of energy for classical systems. Things get more complicated when gravitational fields are present.
\par
First of all there is no fully satisfactory concept of energy in general gravitational backgrounds. In de Sitter space  a Killing energy with support on the space orthogonal to a given observer, $u$, is well-defined through
\be
E(u)\equiv \int d^{n-1}x\,u_\m T^{\m\n}k_\n
\ee
where the energy-momentum tensor is defined by expanding \`a la Abbott-Deser around a background. The lack of global existence of the Killings means that precise statements are only possible outside the corresponding horizons.
In the general situation the situation is even worse, and several definitions (such as the Hawking-Geroch, Penrose, Nester-Witten or Brown-York, \cite{Szabados}) of quasilocal energy exist, none of which is fully satisfactory, and  besides all of them seem difficult to compute in quantum field theory. 
\par
Besides it is  the case in general that
\be
|0_+\rangle \neq|0_-\rangle
\ee
The usual Feynman path integral computes expectation values 
\be
\langle 0_+|{\cal O}|0_-\rangle
\ee
so that some modification is in order to get expectation values such as
\be
\langle 0_-|{\cal O}|0_-\rangle
\ee
One way to do it is the closed time path (CTP) formalism of Schwinger and Keldysh \cite{Schwinger}, but euclidean methods are also available \cite{Vilkovisky}.
\par
The proper approach would be to study the structural stability of the Dyson-Schwinger equations for the whole system.
\par
What we have done in this paper instead is to compute the simplest
 and most naive expression for the energy, namely the effective potential.\\
\bi
\item
As a matter of fact, the formula (\ref{heatkernel}) for the sphere $S_n$ could be directly continued to de Sitter space, given that the Gegenbauer polynomials $C_j^{\frac{n-1}2}$ are defined for all real $z$. Then, the expression:
\begin{equation}
K(\tau;z)=\frac 1{V(S_n)}\sum_j\frac{n-1+2j}{n-1}\,C^\frac{n-1}{2}_j(z)e^{-\tau(m^2l^2+V^{\prime\prime}\left(\bar{\phi}\right)+j(j+n-1))}
\end{equation}
is a natural candidate for the heat kernel in de Sitter as well.\footnote{ It seems plain that the analytic continuation, should it work at all, it not will do it term by term. The eigenvalues are not the same in the sphere as in de Sitter space, not to mention the fact that the sphere is a compact space whereas de Sitter is not. Nevertheless, there is a well-known {\em duality} between compact and non-compact symmetric spaces \cite{Helgason}. Some further caveats on the analytic continuation of the heat kernel
have been made in \cite{Avramidi}. It is true that until the whole sum is performed and then the explicit continuation is made, surprises may appear, so perhaps some wise restrain is called for.}\\

Then we can evaluate the free energy given by formula (\ref{freeenergy}):
\[
W=\frac12\int_0^\infty \frac{d\t}{\t}\,\int d^n x \sqrt{|g|} K\left(\t;x,x\right)=\frac{\textrm{Vol}_{dS}}2\int_0^\infty \frac{d\t}{\t}K(\tau;1)=
\]
\begin{equation}
=\frac{\textrm{Vol}_{dS}}{2V(S_n)}\int_0^\infty \frac{d\t}{\t}\sum_j\frac{n-1+2j}{n-1}C^\frac{n-1}{2}_j(1)e^{-\frac\tau{\mu^2}(m^2+V^{\prime\prime}\left(\bar{\phi}\right)+j(j+n-1)/l^2)}
\end{equation}
where we have redefined the heat kernel in order to get a mass dimension 2 equation. Here $C_j^\frac{n-1}{2}\left(1\right)=\dbinom{j+n-2}{j}$. This expression, which is divergent\footnote{General theorems imply that the trace of the heat kernel must diverge when $\t\rightarrow 0$ as $K\sim \m^n\t^{- n/2}$. This just means that the sum and the integral do not commute.}, is purely real (the $C^\frac{n-1}{2}_l(1)$ are integers), so no imaginary parts appear.\\

\item In the reference \cite{Limic} the spectrum of the laplacian for de Sitter space, $dS_n$, anti de Sitter space $AdS_n$ and euclidean (anti) de Sitter space $EAdS_n$ is computed and the eingenfunctions are constructed as well. The spectrum is identical\footnote{Except for a sign perhaps, depending on the sign chosen for the metric for each space.} for both  $dS_n$ and $AdS_n$ and has got a discrete part (similar to the one corresponding to the sphere)
\[
-L\left(L+n-1\right)/ l^2
\]
where
\[
L=-\left[\frac{n}{2}\right]+1,-\left[\frac{n}{2}\right]+2,\ldots -\left[\frac{n}{2}\right]+j\ldots
\]
and we represent by $\left[z\right]$ the integer part of $z$.
The starting point of the spectrum is actually the only difference between the sphere and both de Sitter and anti de Sitter spaces, as long as the discrete part of the said spectrum is concerned.
In terms of $j\in \mathbb{N}$, for even dimension, $n=2 m$, or else for odd dimension $n=2 m+1$
\[
L=-\frac{-j\left(j-1\right)+m\left(m-1\right) }{4 l^2}
\]

\par
 There is also a continuous piece of the spectrum, which can be written in the form
\[
\frac1{l^2}\left(\Lambda^2+\frac{\left(n-1\right)^2}{4}\right)\textrm{ where }\Lambda\in[0,\infty)
\]
In the case of $EAdS_n$ only the continuous spectrum appears. So the situation is as follows: the two euclidean spaces enjoy only one type of spectrum; discrete in the case of the sphere $S_n$ and continuum in the case of $EAdS_n$; whereas the two manifolds with lorentzian signature ($AdS_n$ and $dS_n$) carry both discrete and continuous spectra.
In all cases the eigenvalues are of course real. 
\par
The eigenfunctions are explicitly known and can be find in the references just quoted. It is enough  for our  purposes though  to point out that they obey a completeness relationship, 
\be
\sum_L Y_L(x)^* Y_L(y)+\int d\Lambda Z_\Lambda(x)^* Z_\Lambda(y)=\d\left(x,y\right)
\ee

\item
Let us nevertheless perform a simple approximation (in the case of the sphere; the other cases are very similar), just to get an idea of the result. We shall explore the high angular momentum region, 
\[\sum_j j^{n-1} e^{-\frac\t{\mu^2}\left(j+n-1\right)j/l^2}\sim  \int_0^\infty d j j^{n-1} e^{-\frac\t{\mu^2l^2}j^2}=
\frac{(\mu l)^n}{2\tau^\frac{n}{2}}\Gamma\left(\frac n2\right)\]
	
\par

We then get in this approximation
\[ W\sim \mu^n l^n\int_{\frac{\mu^2}{\Lambda^2}}^\infty \frac{d\t}{\t^{1+\frac n2}}e^{-\frac{m^2 + V^{\prime\prime}\left(\bar{\phi}\right)}{\m^2}\t} =(m^2l^2+V''(\bar\phi)l^2)^{\frac n2}\,\Gamma\left(-\frac n2,\frac{m^2+V''(\bar\phi)}{\Lambda^2}\right)=
\]
\begin{equation}
=\Bigg\{
\begin{array}{l}
\textrm{odd } n:\ 0\\
\textrm{even }n:\ -\frac{(-1)^{\frac n2}}{(\frac n2)!}(m^2l^2+V''(\bar\phi)l^2)^{\frac n2}\log\frac{\Lambda^2}{m^2+V''(\bar\phi)}+\frac {2\Lambda^n l^n}n+\ldots
\end{array}
\end{equation}

 Here, as  in flat space, the only possible imaginary part comes from the logarithm, that is, when

\[\frac{m^2 + V^{\prime\prime}\left(\bar{\phi}\right)}{\m^2}\leq 0\]

This is in agreement with general theorems \cite{Hadamard} asserting that the only way a non vanishing imaginary part can appear in a manifestly real integral is from the region in which the integral diverges.
\par
On the other hand, his is exactly the situation when spontaneous symmetry breaking occurs in flat space and, as we shall argue in the next paragraph, it is believed to be well understood.

\ei

\newpage
\section{Conclusions}

The effective potential of quantum fields propagating in a constant curvature space, corresponding to a cosmological constant of either sign, has been computed using the heat kernel as our main tool. Most Green's functions that appear obey Polyakov's composition principle, although other possibilities have been examined as well. The general analytic continuation of the sphere
\be
S_n\sim SO(n+1)/SO(n)
\ee
has been considered; we believe this to be physically important, in order to determine whether the purported instability appears only for one sign of the cosmological constant, or for both, in which case it would be possible that the endpoint of the instability would have been flat Minkowski space.

\par
No imaginary part for the effective potential has been obtained except in those cases in which the potential is such that in flat space leads to spontaneous symmetry breaking; that is, when $\pd^2  
V_{eff}(\bar{\phi})<0$
for some range of the argument, like in the famous {\em mexican hat} potentials; and this particular imaginary part is in principle well understood cf. \cite {Weinberg}. It can be shown from first principles 
\cite{Symanzik} that the effective potential $V_{eff}(\bar{\phi})$ corresponds to the expectation value of the energy density in a Fock state $|\Psi\rangle$ which minimizes $\langle \Psi|H|\Psi\rangle$ subject to the constraint    $\langle \Psi |\phi|\Psi\rangle=\bar{\phi}$. This implies that $V_{eff}$ must be real and convex.
\par
What happens for those ranges for which $\pd^2 V_{eff}(\bar{\phi})<0$ is that the state that minimizes the energy (let us call it $|E_0\rangle$) is a quantum superposition of two or more vacuum states, and the configurations for which the expectation value of the field is constant are unstable towards decay into $|E_0\rangle$; the imaginary part $Im\, V_{eff}(\bar{\phi})$ just gives half the decay rate corresponding to this process per unit volume, $\Gamma\left(|\bar{\phi}\rangle\rightarrow |E_0\rangle\right)$.
\par
This is the only imaginary part of the effective potential within the class of models studied in this paper. Our results seem to be compatible with those in \cite{PerezNadal}.

\par
We would like to finish the paper by pointing out an argument 
\footnote{ Related remarks can be found in \cite{Tsamis}}
clarifying when one is to expect instabilities of the background  
field. The fact that the functional integral of a total derivative  
vanishes implies
\[0=\int {\cal D}g_{\m\n}{\cal D}b{\cal D}c{\cal D}\phi\frac{\d}{\d  
g^{\m\n}(x)} e^{i \left(S_{grav}(g_{\m\n})+S_{gf}(g_{\m\n})+S_{gh} 
(b,c,g_{\m\n})+S_m(\phi,g_{\m\n})\right)}\]

When
\[S_{grav}=\frac{1}{2\kappa^2}\int \sqrt{|g|}d^n x R\]
a definition of the composite operators $R_{\m\n}\left(g_{\a\b}\right)$ and $T_{\m\n}\left(g_{\a\b},\phi\right)$ should exist such that the Dyson-Schwinger equation holds:
\[\left\langle \chi\left|\sqrt{|g|}\left(R_{\m\n}-\frac{1}{2}R g_{\m\n}- 
\kappa^2 T_{\m\n}\right)\right|\psi\right\rangle=0\]
where
\[T_{\m\n}\equiv \frac{2}{\sqrt{|g|}}\frac{\d}{\d g^{\m\n}}\left(S_m 
+S_{gf}+S_{gh}\right)\]
and
$|\psi\rangle $ and $|\chi\rangle$ are states that depend on the boundary conditions. Usually they are taken as
$|0^\pm\rangle $.

The trace of the former equation means that
\[
\bar{g}^{\m\n}\left\langle \chi\left|\sqrt{|g|}\left(\frac{2-n}{2}R_{\m\n} -\kappa^2 T_{\m\n}\right) 
\right|\psi\right\rangle=0\]
which means in turn that when the trace of the expectation value of the energy  
momentum is constant, so is the trace of the expectation value of the scalar  
curvature. On the other hand, we insist that both the scalar curvature as well as the  
energy-momentum tensor are composite operators, whose definition is  
somewhat delicate. But this fact also tells us when a nontrivial  
physical effect is at least allowed
 First of all, through the effect of the one-loop gravitational  
counterterms,namely,
\[L_{count}=\int \sqrt{|g|}d^n x \left(c_1 R^2 + c_2 R_{\m\n}^2\right)\]
{\em except} in the renormalization scheme when the finite parts of  
both $c_1$ and $c_2$ are put equal to zero. This changes the contribution of
\[
\frac{\d S_{grav}}{\d g^{\m\n}}
\]
Counterterms are also at the origin of the trace anomaly , i.e.
\[
\bar{g}^{\m\n} \langle T_{\m\n}\rangle\neq   \langle g^{\m\n}T_{\m\n}\rangle
\]
which has got a piece proportional to the beta function of  
the theory, as well as a gravitational piece, which is non-vanishing  
even for conformally invariant theories (i.e., when $\b=0$).\footnote{ They have been classified by
Deser and Schwimmer \cite{Deser} into two types: type A, proportional to the (scale 
invariant) Euler density, and type B
that require introduction of a scale through regularization.}

The conclusion of the analysis is that we do not find any obvious reason why matter effects by themselves
could not destabilize de Sitter space, causing the cosmological constant to decay.
This still looks like an exciting possibility. It remains to find a self-consistent scenario implementing this general idea.
Work on these lines is currently in progress.

\section*{Acknowledgments}
We are grateful to Dani Arteaga, Jaume Garriga, Guillem P\'erez-Nadal, Albert Roura and Enric Verdaguer for many discussions and patient explanations. We also thank Sigurdur Helgason for useful correspondence.
This work has been partially supported by the
European Commission (HPRN-CT-200-00148) and by FPA2003-04597 (DGI del MCyT, Spain) and 
Proyecto HEPHACOS ; P-ESP-00346 (CAM) and CSD 2007 00060(MEC), PAU Consolider. R.V. is supported by a MEC grant, AP2006-01876.

\newpage
\appendix

\section{Taxonomy of constant curvature spaces.}
\FIGURE{
\centerline{
\psfig{file=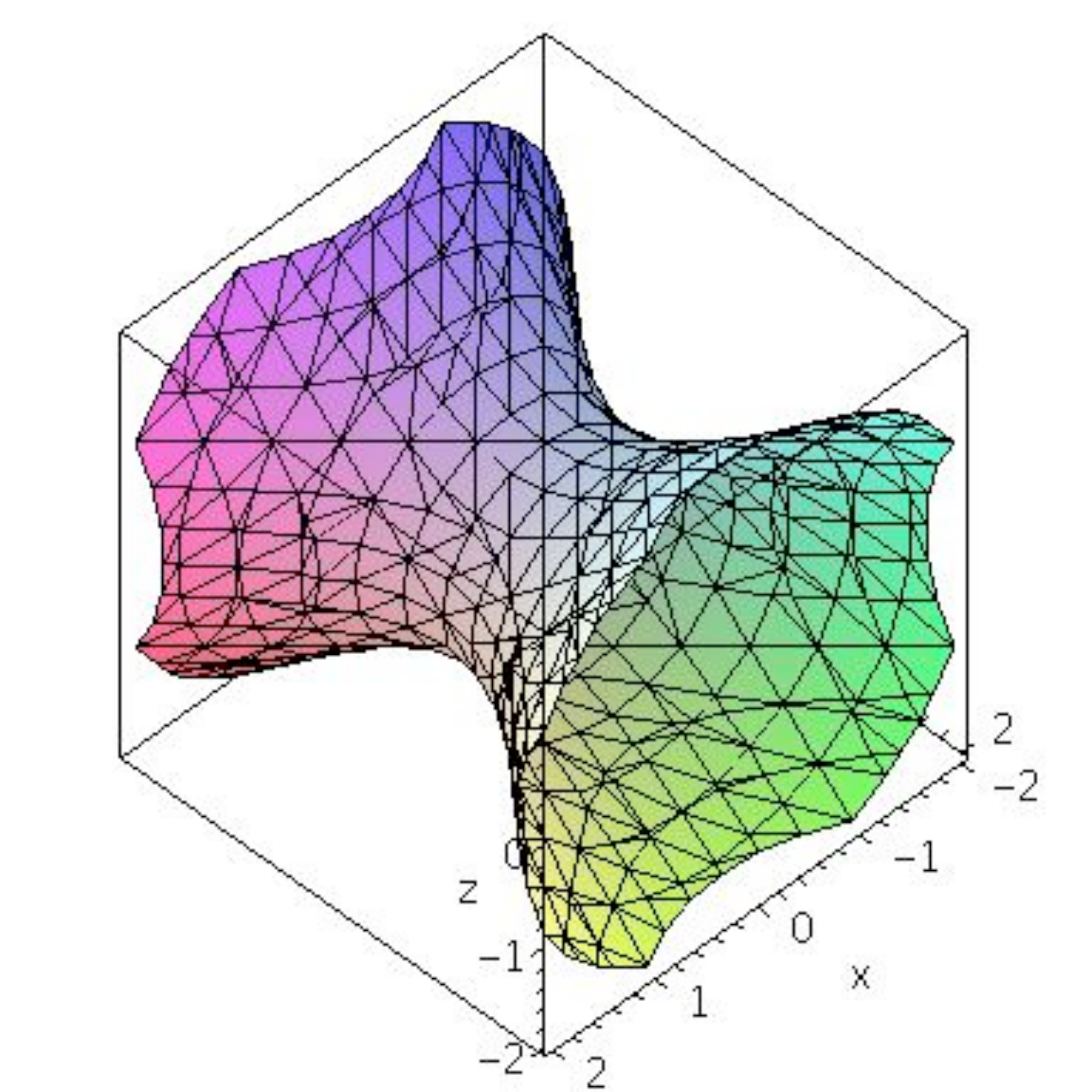,width=9cm}
\caption{ A pictorial representation of Anti de Sitter ($X_0^2+X_1^2=l^2+\vec{X}^2$ in $\mathbb{R}^{n+1}_{n-1}$).
}}
}
The real sections of the complex sphere can be treated in an unified way. Let us choose coordinates in the embedding space in such a way that in the defining equation we have
\be\label{defining}
X^2=\sum_{A=0}^n \e_A X_A^2\equiv \eta_{AB}dX^A dX^B =\pm l^2
\ee
on a flat space with metric $ds^2= \eta_{AB}dX^A dX^B$. If we change in an arbitrary manifold $g_{AB}\rightarrow -g_{AB}$, then both Christoffels and Riemann tensor remain invariant, but the scalar curvature flips sign $R\rightarrow -R$. We can furthermore group together times and spaces, in such a way that
\be
\eta_{AB}=(1^t,(-1)^s)
\ee
If we call $n+1\equiv t+s$, then this ambient space is Wolf's $\mathbb{R}^{n+1}_s$ where the subindex indicates the number of {\em spaces}.
\par
 The standard  nomenclature in Wolf's book \cite{Wolf} is
\[S^n_s: X\in\mathbb{R}^{n+1}_s, \,X^2=l^2\]
\begin{equation}\label{wolf}
H^n_s: X\in\mathbb{R}^{n+1}_{s+1}, \,X^2=-l^2 
\end{equation}

\FIGURE{
\centerline{
\psfig{file=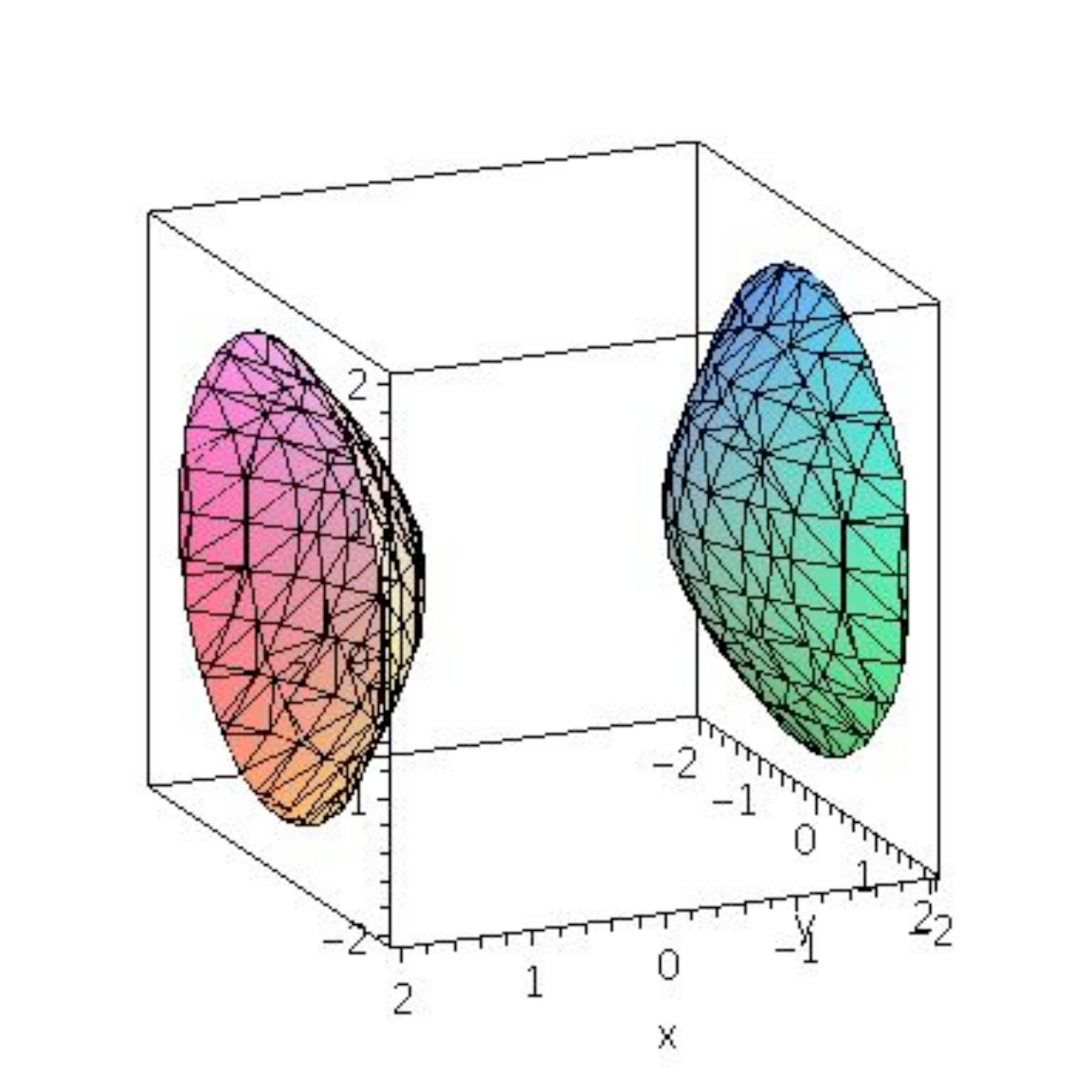,width=9cm}
\caption{ A pictorial representation of Euclidean Anti de Sitter (or Euclidean de Sitter) ($X_0^2=l^2+\vec{X}^2$ in $\mathbb{R}^{n+1}_n$).
}}
\label{2}}

The curvature scalar is given by:
\be
R=\pm \frac{n(n-1)}{l^2}
\ee
and

\bea
&&R_{\m\n}=\pm\frac{n-1}{l^2}g_{\m\n}\nonumber\\
&&R_{\m\n\rho\sigma}=\pm\frac{1}{l^2}\left(g_{\m\rho}g_{\n\sigma}-g_{\m\rho}g_{\n\sigma}\right)
\eea
Please note that the curvature only depends on the sign on the second member, and not on
the signs $\e_A$ themselves.\\

It is clear, on the other hand, that the isometry group of the corresponding manifold is
one of the real forms of the complex algebra $SO(n+1)$. The Killing vector fields are
explicitly given (no sum in the definition) by
\be
L_{AB}\equiv \epsilon_{A}X^{A}\pd_{B}-\epsilon_{B}X^{B}\pd_{A}\equiv X_{A}\pd_{B}-X_{B}\pd_{A}
\ee
The square of the corresponding Killing vector is
\be
L^2= \e_B X_A^2+\e_A X_B^2
\ee

\FIGURE{
\centerline{
\psfig{file=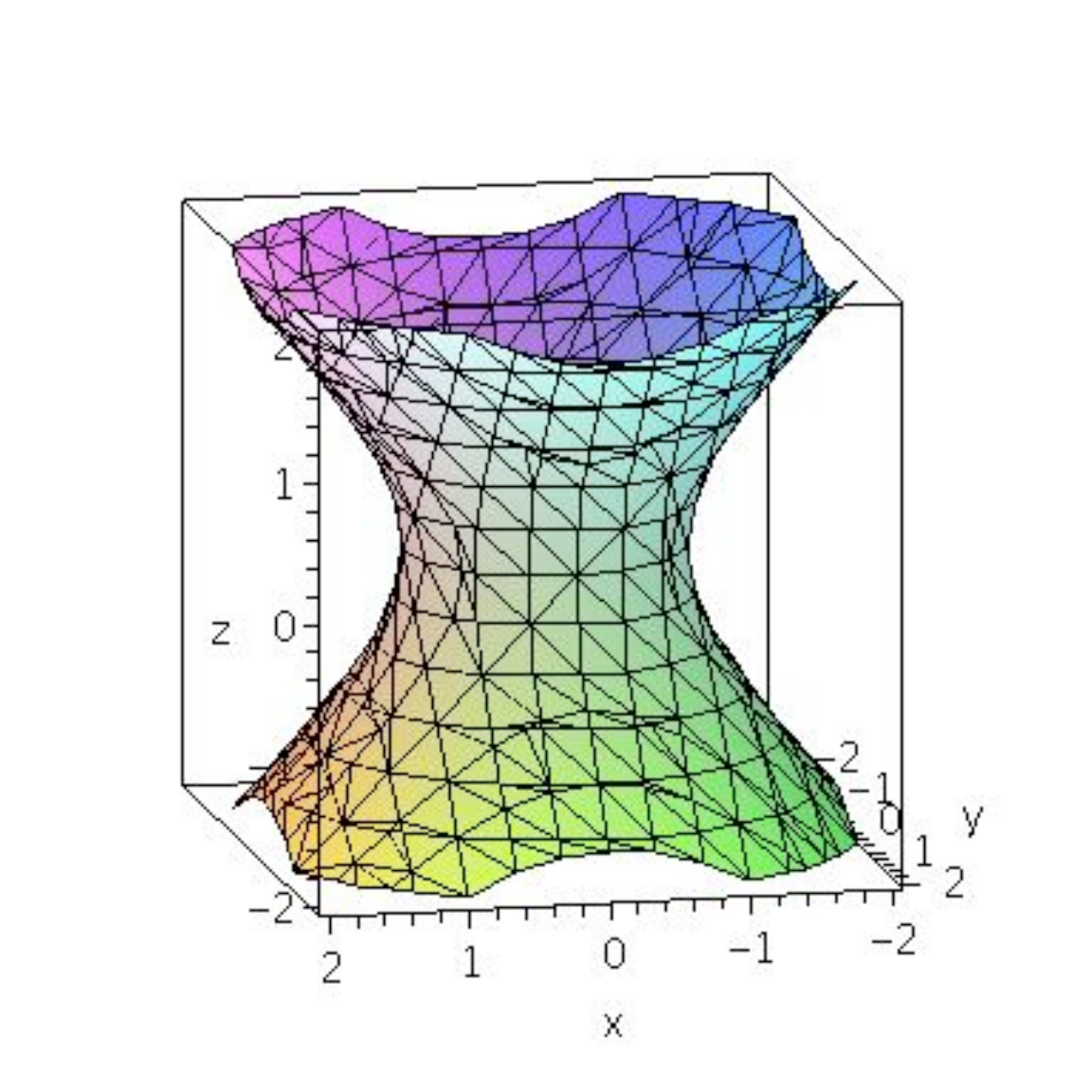,width=9cm}
\caption{ A pictorial representation of de Sitter ($X_0^2=-l^2+\vec{X}^2$) in $\mathbb{R}^{n+1}_n$).
}}
\label{3}}

Our interest is concentrated on the euclidean and minkowskian cases:
\begin{itemize}
\item The sphere $S_n\equiv S^n_0\sim H^n_n$ is defined by $\vec X^2=l^2$, with isometry group $SO(n+1)$.
\item The euclidean Anti de Sitter (or euclidean de Sitter) $EAdS_n\equiv S^n_n\sim H^n_0$ is defined by $(X^0)^2-\vec X^2=l^2$, with isometry group $SO(1,n)$.
\item The de Sitter space $dS_n\equiv H^n_{n-1}\sim S^n_1$ is defined by $(X^0)^2-\vec X^2=-l^2$, with isometry group $SO(1,n)$. In our 
conventions de Sitter has negative curvature, but positive cosmological constant.
\item The Anti de Sitter space $AdS_n\equiv S^n_{n-1}\equiv H^n_{1}$ is defined by $(X^0)^2+(X^1)^2-\vec X^2=l^2$, with isometry group $SO(2,n-1)$. For us $AdS_n$ has positive curvature and negative cosmological constant.
\end{itemize}

\subsection{Global coordinates}\label{zeta}
A very useful coordinate chart for these spaces is the one called \textit{global} coordinates, wich nevertheless do not cover the full space in any case:
\begin{equation}
(X^A)=l\,\left(\cosh\tau\,\vec u_{t}(\Omega),\sinh\tau\,\vec n_{s}(\Omega')\right)
\end{equation}
where $\vec u$ and $\vec n$ are unit vectors of both $t-1$ and $s-1$ dimensional spheres. This is for $S^n_s$ spaces. For $H^n_s$ spaces is simply:
\begin{equation}
(X^A)=l\,\left(\sinh\tau\,\vec u_{t-1}(\Omega),\cosh\tau\,\vec n_{s+1}(\Omega')\right)
\end{equation}
Our convention for a unit vector of a ($n-1$)-dimensional sphere is:
\begin{equation}
\vec u_n(\Omega)=(\cos\theta_1,\sin\theta_1\cos\theta_2,\ldots,\sin\theta_1\ldots\sin\theta_{n-1})
\end{equation}
so that our convention for the ``north pole'' is:
\begin{equation}
S^n_s\textrm{: }\ N=(l,0,\ldots)\textrm{ ; }H^n_s\textrm{: }\ N=(\underbrace{0,\ldots}_{t-1},l,0,\ldots) 
\end{equation}

The invariant distance, that we call $z$, is defined as $z(X,Y)=\pm\frac{X\cdot Y}{l^2}$, where the sign is chosen to make $z(X,X)=1$ in every space. In our cases of interest:
\begin{itemize}
\item Sphere: $X=l\,\vec u_n(\Omega)$, $z=\cos\theta_1$
\item Euclidean Anti de Sitter: $X=l(\cosh\tau,\sinh\tau\vec u_{n-1}(\Omega))$, $z=\cosh\tau$
\item de Sitter: $X=l(\sinh\tau,\cosh\tau\,\vec u_{n-1}(\Omega))$, $z=\cosh\tau\cos\theta_1$
\item Anti de Sitter: $X=l(\cosh\tau\cos\theta,\cosh\tau\sin\theta,\sinh\tau\vec u_{n-2}(\Omega'))$, $z=\cosh\tau\cos\theta$
\end{itemize}

\subsection{Projective coordinates}
We shall further assume that $\e_k=\pm 1$, that is, the choosen coordinate has the same sign for the metric as the second member in (\ref{wolf}). We then define the south pole (i.e. $X^k=-l$) stereographic projection for $\m \neq k$, as
\be
x^\m_S\equiv \frac{2l}{X^k+l} X^\m\equiv \frac{X^\m}{\Omega_S}
\ee
The equation of the surface then leads to 
\begin{equation}
X^k=l(2\Omega_S-1)\ \textrm{;}\ \Omega_S=\frac{1}{1\pm\frac{x_S^2}{4l^2}}\textrm{ ; }x_S^2\equiv\sum_{\m \neq k}\e_\m \left(x_S^\m\right)^2
\end{equation}

The metric in these coordinates is conformally flat:
\be
ds^2= \Omega_S^2 \eta_{\m\n}dx_S^\m dx_S^\n
\ee

We could have done projection from the North pole (for that we need that
$X^k\neq l$). Uniqueness of the definition of $X^k$ needs
\be
\Omega_N+\Omega_S=1
\ee
and uniqueness of the definition of $X^\m$
\be
x^\m_N=\frac{\Omega_S}{\Omega_N}x^\m_S=\pm\frac{4 l^2}{x_S^2} x^\m_S
\ee

The antipodal $\mathbb{Z}_2$ map $X^A\rightarrow -X^A$
is equivalent to a change of the reference pole  in stereographic coordinates

\be
x^\m_N\leftrightarrow x^\m_S
\ee

\subsection{Poincar\'e coordinates}
A generalization of Poincar\'e's metric for the half-plane can easily be obtained by 
introducing the horospheric coordinates.
It will always be assumed that $\e_0=+1$, that is that $X^0$ is a time, and also 
that $\e_n=-1$, that is $X^n$ is a space, in our conventions. Otherwise (like in 
the all-important case of the sphere $S_n$) it it not possible to construct these coordinates.

\bea
&&\frac{l}{z}\equiv X^{-}=X^n-X^0\nonumber\\
&&y^{i}\equiv z X^{i}
\eea
The promised generalization of the Poincar\'e metric is:
\be
ds^2 = \frac{\sum_1^{n-1} \epsilon_{i} dy_{i}^2 \mp l^2 dz^2}{z^2}
\ee
where the sign is the opposite to the one defined in (\ref{wolf}), 
and the surfaces of constant $z$ are sometimes called {\em horospheres}. This form of the metric 
is conformally flat  in a manifest way.

\bi
\item In {\bf de Sitter 
space}, $dS_n$, $z$ is a timelike coordinate, and its metric reads
\be
ds_{dS_n}^2=\frac{-\sum^{n-1}\delta_{ij}dy^i dy^j+ l^2 dz^2}{z^2}
\ee
The square of the Killing vectors $M_{0A}$ (candidates to be timelike)
are
\be
M_{0A}^2=X_0^2-X_A^2=\sum_{B\neq A}X_B^2-l^2
\ee
so they are timelike only outside the {\em horizon} defined as
\be
H_{0A}\equiv \sum_{B\neq A}X_B^2=l^2
\ee
For example, the horizon corresponding to $H_{0n}$ is
\be
\sum y_i^2= l^2 z^2
\ee
This means that de Sitter space, $dS_n$ is not globally static.

\item What one would want to call {\bf Euclidean anti de Sitter}, 
$EAdS_n$, has got all its coordinates spacelike, and {\em positive} curvature.
To be specific
\be
ds_{EAdS_n}^2=\frac{-\sum^{n-1}\delta_{ij}dy^i dy^j- l^2 dz^2}{z^2}
\ee
\item Finally, when the metric is given by
\be
ds_{AdS_n}^2=\frac{\sum^{n-1}\eta_{ij}dy^i dy^j- l^2 dz^2}{z^2}
\ee
(where as usual, $\eta_{ij}\equiv \textrm{diag}(1,-1^{n-2})$) this is the {\bf Anti de Sitter}, $AdS_n$. In this case there is a globally defined timelike Killing vector field, namely $M_{01}$
\be
M_{01}^2=X_0^2+ X_1^2= l^2+ \sum_{A> 1} X_A^2
\ee
that is everywhere positive. This means that Anti de Sitter space is globally static, as opposed to de Sitter.
\ei

\subsection{Conformal Invariance}

Let us be very explicit with the definition of Poincar\'e coordinates:
Let us denote
\be
x^2\equiv y^2\mp l^2 z^2\equiv \sum \e_i y_i^2\mp l^2 z^2
\ee
Then 
\bea
&&X^0=\frac{l^2-x^2}{2 l z }\nonumber\\
&&X^n=-\frac{l^2+x^2}{2 l z }\nonumber\\
&&X^i= \frac{y^i}{z}\,(i=1\ldots n-1)
\eea
This is a legitimate change of coordinates as long as we keep the radius $l$ itself as one of the coordinates.

Conversely,
\bea
&&y^i=\frac{X^i}{X^0 - X^n}l\nonumber\\
&&z=\frac{l}{X^0-X^n}\nonumber\\
&&l^2=\mp\left(X_0^2-X_n^2+\e_i X_i^2\right)
\eea
Some useful formulas:
\bea
&&\frac{\pd }{\pd X_0}=-\frac{z}{l}y^i \pd_i-\frac{z^2}{l}\pd_z \mp \frac{l^2- x^2}{l z}\pd_{l^2}\nonumber\\
&&\frac{\pd}{\pd X_n}=\frac{z}{l}y^i \pd_i+\frac{z^2}{l}\pd_z \mp \frac{l^2+ x^2}{l z}\pd_{l^2}\nonumber\\
&&\frac{\pd}{\pd X_i}=z\pd_i\mp 2 \frac{\e_i y_i}{z}\pd_{l^2}
\eea
The full isometry group is some noncompact form of $SO(n+1)$. In Poincare coordinates, there is a $ISO(n-1)$ manifest isometry group not involving the horographic coordinate. It will be important for us to understand all isometries in Poincar\'e coordinates.
Let us work out the non-explicit generators:
\[
L_{0n}\equiv X^0\pd_n+X_n\pd_0=y^i \pd_i+z\pd_z
\]
\[
L_{0i}=X^0\pd_i-\e_i X_i \pd_0=\sum_j\frac{\left(l^2-x^2\right)\d_{ij}+2\e_i y_i y_j}{2l}\pd_j+\e_i y^i \frac{z}{l}\pd_z
\]
\[
L_{ni}=-X^n\pd_i-\e_i X_i \pd_n=\sum_j\frac{\left(l^2+x^2\right)\d_{ij}-2\e_i y_i y_j}{2l}\pd_j-\e_i y^i \frac{z}{l}\pd_z
\]

Translations of the $y^i$ correspond to the combination: 
\be
k_i\equiv l\frac{\pd}{\pd y^i}=-\left(L_{ni}+L_{oi}\right)
\ee

All spaces we are considering in this paper, which in Poincar\'e coordinates enjoy the metric
\be
ds^2=\frac{\sum^{i=n-1}_{i=1} \e_i dy_i^2 \mp l^2 dz^2}{z^2}
\ee

are obviously {\em scale invariant}
\bea
&&y_i\rightarrow \l\,y_i\nonumber\\
&&z\rightarrow \l\,z
\eea
This corresponds in Weierstrass coordinates to the Lorentz transformation in the plane $\left(X^0 X^n\right)$
\bea
&&\left(X^\prime\right)^0=\frac{\left(\l^2+1\right)X^0+\left(\l^2-1\right)X^n}{2 \l}\nonumber\\
&&\left(X^\prime\right)^n=\frac{\left(\l^2-1\right)X^0+\left(\l^2+1\right)X^n}{2 \l}
\eea
id est, 
\bea
&&X^-\rightarrow \l X^-\nonumber\\
&&X^+\rightarrow \frac{X^+}{\l}
\eea
(This ought to be more or less obvious  already from the previous formula for the generator $L_{0n}$).
Not only that, but also they are invariant under {\em inversions}, id est,
\bea
 &&y_i\rightarrow \frac{y_i}{\sum \e_i y_i^2 \mp l^2 z^2}\nonumber\\
&&z\rightarrow \frac{z}{\sum \e_i y_i^2 \mp l^2 z^2}
\eea
Inversions in Weierstrass coordinates look even simpler; just  exchange the two light-cone coordinates in the aforementioned plane $\left(X^0 X^n\right)$:
\be
X^+\leftrightarrow X^-
\ee
The remaining isometries are the somewhat nasty combinations
\be
L_{0i}-L_{ni}=\sum_j\frac{\left(-x^2\right)\d_{ij}+2\e_i y_i y_j}{l}\pd_j+2\e_i y^i \frac{z}{l}\pd_z
\ee
\par
We are now in a position to study the little group $H$ of a given point (which can always be rotated to
\be
P\equiv\left(\vec{y}=\vec{0},z=1\right)
\ee
 We know that then the space will be isomorphic to $SO(n+1)/H$. The translational isometries  must be generated by the $n$ generators
\bea
&&L_{ni}+L_{0i}\nonumber\\
&&L_{0n}
\eea
It seems then that 
\bea
&&H^+=\{L_{ij},L_{ni}\}\nonumber\\
&&H^- =\{L_{ij},L_{0i}\}
\eea
 The number of not compact generators is equal to the number of times in the coordinates $y^i$ in the + case, and the number of times plus one in the minus case. 
This seems to imply that
\bea
&&AdS_n=SO(2,n-1)/SO(1,n-1)\nonumber\\
&&EAdS_n=SO(1,n)/SO(n)\nonumber\\
&&dS_n=SO(1,n)/SO(1,n-1)\nonumber\\
&&EdS_n=SO(n,1)/SO(n)
\eea
\par
Euclidean anti de Sitter $EAdS_n$ is just de Sitter $dS_n$ with imaginary radius. Euclidean de Sitter $EdS_n$ is Euclidean anti de Sitter $dS_n$ with negative ambient metric.

\newpage

\section{Conformal structure}
\bi
\item ${\bf dS_n}$
From the global coordinates in de Sitter (cf. \ref{zeta}), we can define $\cos T=\frac1{\cosh\tau}$ where $- \pi/2\leq T \leq \pi/2$ so it yields
\be
ds^2= \frac{l^2}{\cos^2 \,T}\left(dT^2- d\Omega_{n-1}^2\right)
\ee
which is conformal to a piece of $\mathbb{R}\times S_{n-1}$, which is the Einstein static universe
to study
conformal structure. The piece is a slab in the timelike direction, but otherwise including the full three-sphere at each time.
The fact that conformal infinity is spacelike means that there are both particle and event horizons.

\FIGURE{
\centerline{
\psfig{file=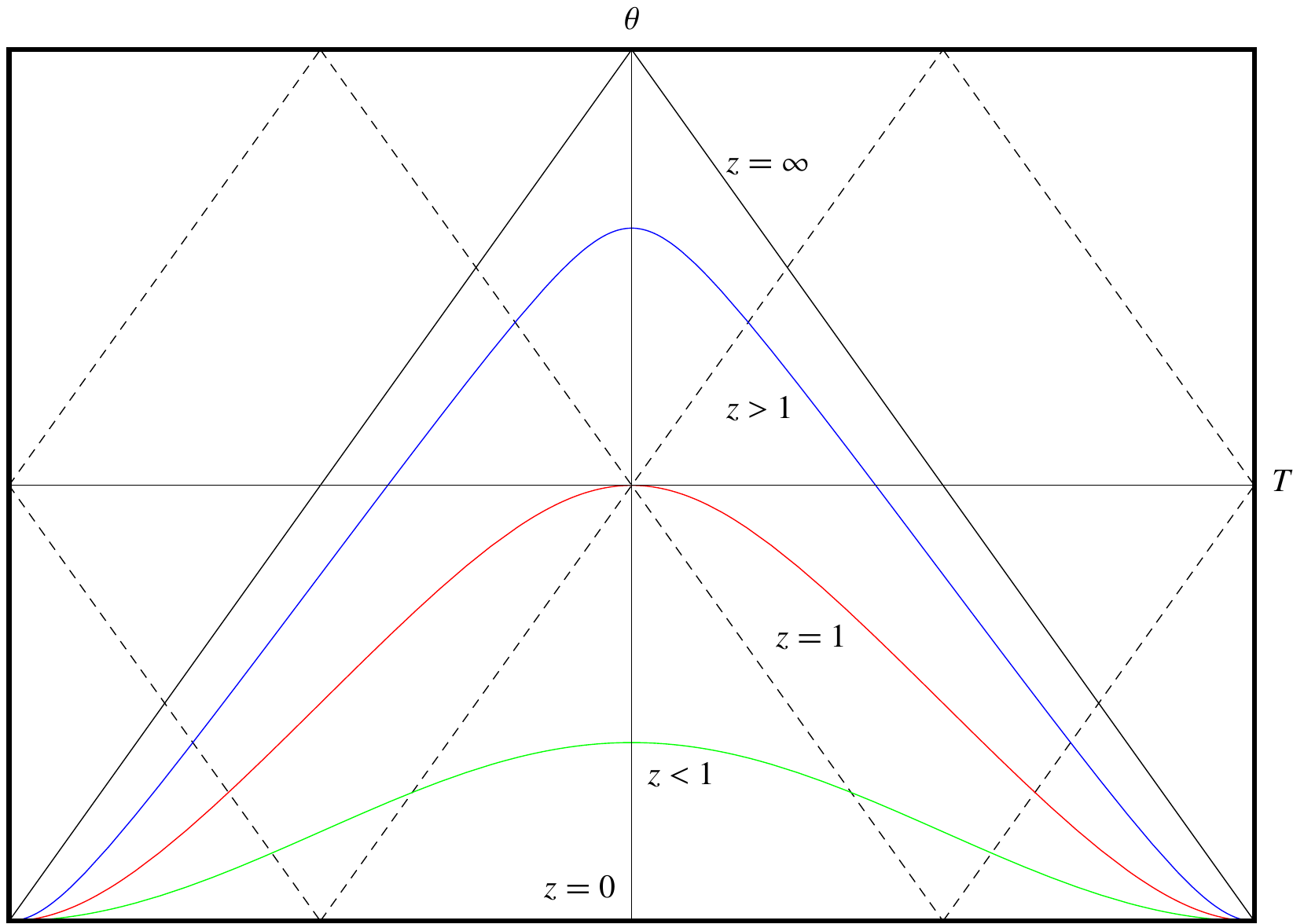,width=13cm}
\caption{ Conformal structure of $dS_n$. The coloured lines are $z=$const. surfaces in Poincar\'e coordinates.}}
\label{10}}

\item ${\bf AdS_n}$
The same change of coordinates from the global chart can be used, $\cos\rho=\frac1{\cosh\tau}$, where $\rho\in(0,\pi/2)$. The space is again conformal to a piece of half Einstein' s static universe:
\be
ds^2= \frac{l^2}{\cos^2\rho}\left(d\theta^2-d\rho^2-\sin^2\rho d\Omega_{n-2}^2\right)=\frac{l^2}{\cos^2\rho}\left(d\theta^2-d\Omega_{n-1}^2\right)
\ee

If we want to eliminate the closed timelike lines, one can consider the covering space $-\infty\leq \theta\leq \infty$.
The slab of $\mathbb{R}\times S_{n-1}$ to which $AdS_n$ is conformal to includes now the full timelike direction, but only an hemisphere
at each particular time. Null and spacelike infinity can be considered as the timelike surfaces $\rho=0$ and $\rho=\pi/2$. This implies that there are no Cauchy surfaces.

\ei

\section{What portion of Weiersstrass coordinates do Poincar\'e coordinates cover?}
\bi
\item ${\bf dS_n}$

If we call $u_n$ the n-th component of the unit vector $\vec{u}$, then there is a critical value of the parameter $\t$ such that
\be
\tanh\,\t(u)=u_n(\Omega)
\ee
which is such that
\be
\t < \t(u)\Rightarrow z > 0
\ee
and
\be
z\rightarrow\pm\infty\Leftrightarrow \t\rightarrow \t(n)^{\mp}
\ee

\FIGURE{
\centerline{
\psfig{file=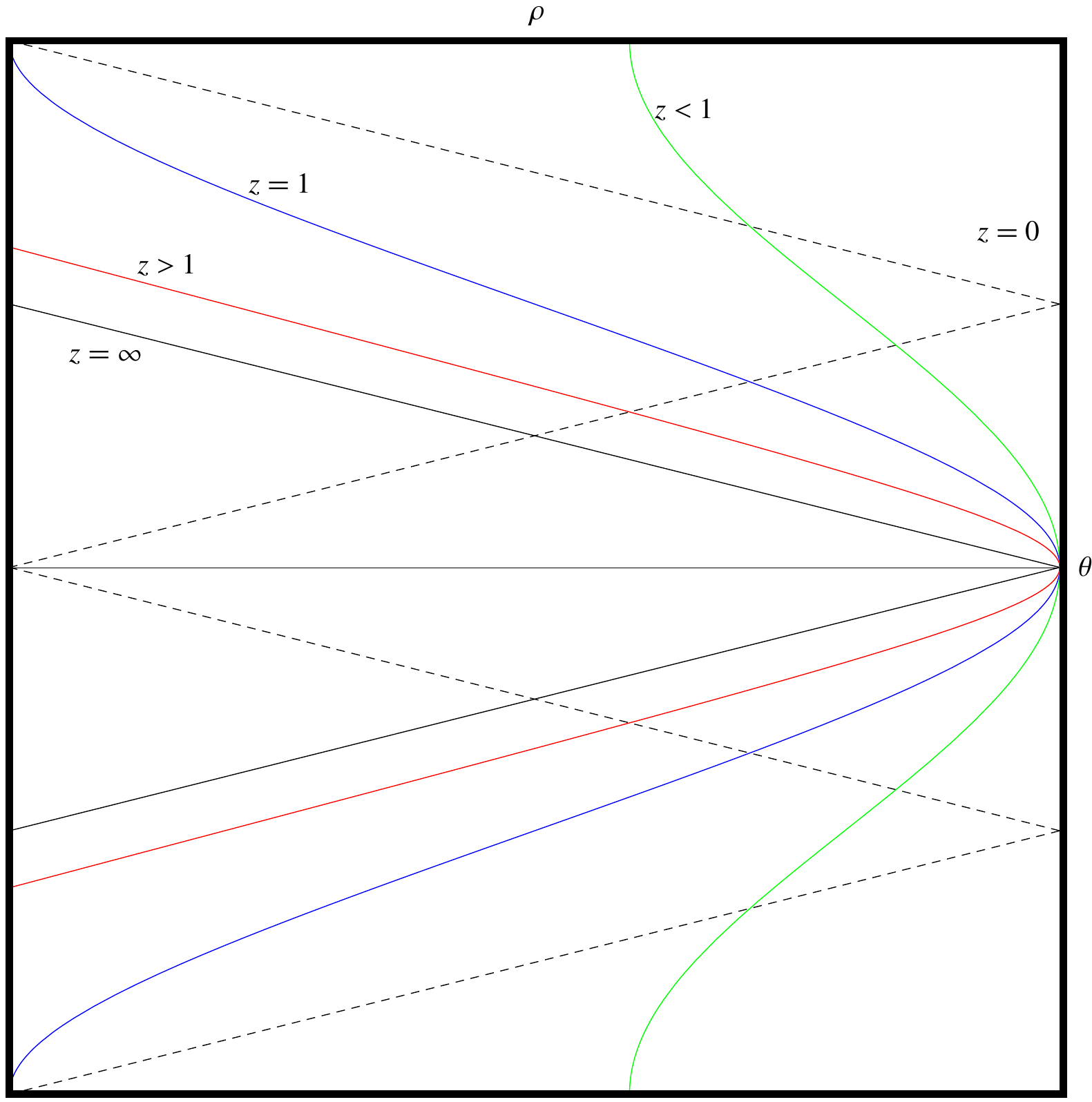,width=13cm}
\caption{ Conformal structure of $AdS_n$. The coloured lines are $z=$const. surfaces in Poincar\'e coordinates.
}}
}

This means that at any given value of $\t$ only those points on the sphere that obey
\be
u_n(\Omega)\geq \tanh\,\t
\ee
can be represented in Poincar\'e coordinates. For example, when $\t=\infty$, that is $T=\pi/2$, $\tanh\,\t=1$, so that
only the North pole ($n=1$) can be covered. At the other extreme, when, $\t=-\infty$, that is $T=-\pi/2$, $\tanh\,\t=-1$, we can cover the full sphere.
\par

On the other hand, it is clear that
\be
z\rightarrow 0^{\pm}\Leftrightarrow \t\rightarrow \mp\infty
\ee

There is a discontinuity at $\t(n)$ which depends on the point in de Sitter space.

\item ${\bf AdS_n}$

As in the previous case, it is clear that the region $1/z=0$ corresponds to 
\be\label{ads-h}
u_{n-1}(\Omega)\sin\rho=\cos\theta
\ee
and the region $z>0$ to
\be
u_{n-1}(\Omega)\sin\rho>\cos\theta
\ee
The region
\be
z=0
\ee
is dubbed the {\em boundary} (of the Poincar\'e patch) of $AdS$ and corresponds to 
\be
\rho=\pi/2
\ee
\par

Finally 
\be
z=\infty
\ee
is usually called the {\em horizon} and corresponds to (\ref{ads-h})

\ei
\newpage
\section{Spherical harmonics}
\bi
\item {\bf The n-dimensional sphere}. The simplest way of getting eigenfunctions of the Laplace operator in the sphere is Helgason's
(confer \cite{Helgason}).
Consider the following harmonic polynomial in $\mathbb{R}^{n+1}$
\be
f_{a,\l}\equiv \left(\vec{a}.\vec{x}\right)^\l
\ee
with $\vec{a}\in\mathbb{C},\,\vec{a}^2=0$. 
\par
Now we know that the full laplacian in $\mathbb{R}^{n+1}$ is
\be
\Delta_{\mathbb{R}^{n+1}}=\frac{\pd^2}{\pd r^2}+\frac{n}{r}\frac{\pd}{\pd r}+\frac{1}{r^2}\Delta_{S_n}
\ee
This yields

\be
\Delta_{\mathbb{R}^{n+1}}f_{a,\l}=0=\frac{\l^2+(n-1)\l}{r^2}f_{a,\l}+\frac{1}{r^2}\Delta_{S_n}f_{a,\l}
\ee
so that the eigenvalues of the Laplacian in the sphere $S_n$ are
\be
-\l(\l+n-1)
\ee
It is more or less equivalent to start from traceless homogeneous polynomials
\be
P\equiv\sum P_{\left(i_1\ldots i_k\right)}x^{i_1}\ldots x^{i_k}
\ee
The number of such animals is the number of symmetric polynomials in n variables of degree $\l$ minus the number of symmetric polynomials of degree $\l-2$:
\be
d(\l)=\binom{\l+n-1}{\l}-\binom{\l+n-3}{\l-2}=\frac{\left(n+2\l-2\right)\left(\l+n-3\right)!}{\l!\left(n-2\right)!}
\ee

\item
If we represent by $\m$ an appropiate collection of indices, then we first build harmonic polynomials such that
\be
\int_{S_n} d\Omega h^*_{\l^\prime\m^\prime}h_{\l\m}=\d_{\l \l^\prime}\d_{\m\m^\prime}r^{\l+\l^\prime}
\ee
The hyperspherical harmonics are then defined by
\be
h_{\l\m}\equiv r^\l Y_{\l\m}
\ee
and are normalized in such a way that
\be
\int_{S_n} d\Omega Y^*_{\l^\prime\m^\prime}Y_{\l\m}=\d_{\l\l^\prime}\d_{\m\m^\prime}
\ee

\item
Gegenbauer polynomials are generalizations of Legendre polynomials, in the sense that
\be
\frac{1}{|\vec{x}-\vec{x^\prime}|^{n-2}}=\frac{1}{r_>^{n-2}\left(1+\left(\frac{r_<}{r>}\right)^2-2\left(\frac{r_<}{r_>}\right)\hat{x}.\hat{x}^\prime\right)^\frac{n-2}{2}}=\frac{1}{r^{n-2}_>}\sum_{\l=0}^\infty\left(\frac{r_<}{r_>}\right)^\l C_\l^\frac{n-2}{2}\left(\hat{x}.\hat{x}^\prime\right)
\ee
Let us now prove the  sum rule for hyperspherical harmonics. For concreteness, let us assume that
\bea
&&r\equiv |\vec{x}_<|\nonumber\\
&&r^\prime\equiv |\vec{x}_>|
\eea
Then it is a fact of life that
\be
\Delta\frac{1}{|\vec{x}-\vec{x}^\prime|^{n-2}}=0=\sum_{\l=0}^\infty \frac{1}{\left(r^\prime\right)^{\l+n-2}}\Delta\left(r^\l C_\l^\frac{n-2}{2}\left(\hat{x}.\hat{x}^\prime\right)\right)
\ee
Imposing term by term vanishing leads to
\be
\left(\frac{1}{r^{n-1}}\frac{\pd}{\pd r}r^{n-1}\frac{\pd}{\pd r}-\frac{1}{r^2}\Delta_{S^{n-1}}\right)\left(r^\l C_\l^\frac{n-2}{2}\left(\hat{x}.\hat{x}^\prime\right)\right)=0
\ee
which conveys the fact that
\be
\Delta_{S_{n-1}} C_\l^\frac{n-2}{2}\left(\hat{x}.\hat{x}^\prime\right)=-\l\left(\l+n-2\right) C_\l^\frac{n-2}{2}\left(\hat{x}.\hat{x}^\prime\right)
\ee
Since the hyperspherical harmonics are by assumption a complete set of eigenfunctions,
\be
C_\l^\frac{n-2}{2}\left(\hat{x}.\hat{x}^\prime\right)
=\sum_\m a_{\l\m}\left(\vec{x}^\prime\right)Y_{\l\m}\left(\hat{x}\right)
\ee
where
\be
a_{\l\m}\left(\vec{x}^\prime\right)=\int_{\hat{x}} C_\l^\frac{n-2}{2}\left(\hat{x}.\hat{x}^\prime\right)Y^*_{\l\m}\left(\hat{x}\right)=\frac{2 (n-2)\pi^{n/2}}{\Gamma(n/2)\left(2\l+n-2\right)}Y^*_{\l\m}\left(\hat{x}^\prime\right)
\ee

This is related to the degeneracy $d(\l)$ of hyperspherical harmonics in the following way. Choosing $\hat{x}=\hat{x}^\prime$, the sum rule leads to
\be
C_\l^\frac{n-2}{2}\left(1\right)
=K_\l\sum_\m Y^*_{\l\m}\left(\vec{x}^\prime\right)Y_{\l\m}\left(\hat{x}\right)
\ee
Integrating now over the unit sphere
\be
C_\l^\frac{n-2}{2}\left(1\right)V(S_{n-1})
=K_\l\sum_\m 1=K_\l d(\l)
\ee

The result is
\be
d(\l)=\frac{\left(n+2\l-2\right)\left(\l+n-3\right)!}{\l!\left(n-2\right)!}
\ee

\item Let us now become more specific and perform some computations in gory detail. 
The metric on $S_{n}$ is
\be
ds_{n}^2=d\theta_{n}^2+ \sin^2\,\theta_{n}d\theta^2_{n-1}+\ldots+\sin^2\,\theta_{n}\,\sin^2\,\theta_{n-1}\ldots \sin^2\,\theta_2 d\theta_1^2
\ee
id est, in a recurrent form
\bea
&&ds_1^2=d\theta_1^2\nonumber\\
&&ds_n^2=d\theta_n^2+\sin^2\theta_n\,ds^2_{n-1}
\eea
This corresponds to polar coordinates in $\mathbb{R}^n$
\bea
&&X_{n+1}=\cos\,\theta_{n}\nonumber\\
&&X_{n}=\sin\,\theta_{n}\,\cos\,\theta_{n-1}\nonumber\\
&&\ldots\nonumber\\
&&X_2=\sin\,\theta_{n}\,\sin\,\theta_{n-1}\ldots \cos\,\theta_1\nonumber\\
&&X_1=\sin\,\theta_{n}\,\sin\,\theta_{n-1}\ldots \sin\,\theta_1
\eea

Spherical harmonics have been constructed quite explicitly by Higuchi \cite{Higuchi}, are such that
\be
\Delta_n Y_{j_n\ldots j_1}(\theta_n\ldots\theta_1)=-j_n(j_n+n-1)Y_{j_n\ldots j_1}(\theta_n\ldots\theta_1)
\ee 
We shall explicitly write down the laplacian in a moment.
They are orhonormal with respect to the induced riemannian measure
\be
d\Omega_n\equiv \sqrt{|g|}d\theta_1\wedge\ldots d\theta_n=d\theta_1\ldots d\theta_n \sin^{n-1}\theta_n\,\sin^{n-2}\,\theta_{n-1}\ldots \sin\,\theta_2 
\ee
The laplacian is easily found to be
\bea
&&\Delta_{S_n}=\left(\frac{\pd^2}{\pd \theta_{n}^2}+\left(n-1\right)\cot\,\theta_n\frac{\pd}{\pd \theta_n}\right)+\frac{1}{\sin^2\theta_n}\left(\frac{\pd^2}{\pd \theta_{n-1}^2}+\left(n-2\right)\cot\,\theta_{n-1}\frac{\pd}{\pd \theta_{n-1}}\right)+\ldots\nonumber\\
&&+\frac{1}{\sin^2 \theta_n\,\sin^2 \theta_{n-1}\ldots \sin^2\theta_2}\frac{\pd^2}{\pd\theta_1^2}
\eea
Another useful recurrence
\be
d\Omega_n=\sin^{n-1}\theta_n d\theta_n d\Omega_{n-1}
\ee
and
\be
V(S_{n-1})=\int d\Omega_{n-1}=\frac{2\pi^{n/2}}{\Gamma(n/2)}
\ee
To be specific,
\be
\int d\Omega_n Y_{j_n\ldots j_1}(\theta_n\ldots\theta_1)Y^*_{j^{\prime}_n\ldots j^{\prime}_1}(\theta_n\ldots\theta_1)=\d_{j_n,j^{\prime}_n}\ldots\d_{j_n,j^{\prime}_n}
\ee

\item It is obvious that any function on the sphere can be expanded
\bea
&&f(\Omega)=\sum_{j_n\ldots j_1} C_{j_n\ldots j_1} Y_{j_n\ldots j_1}(\theta_n\ldots\theta_1)=\nonumber\\
&&\sum_{j_n\ldots j_1}  \int d\Omega^\prime Y^*_{j_n\ldots j_1}(\theta^\prime_n\ldots\theta^\prime_1)f(\theta^\prime_n\ldots\theta^\prime_1) Y_{j_n\ldots j_1}(\theta_n\ldots\theta_1)\nonumber
\eea

which means
\be
\sum_{j_n\ldots j_1} Y^*_{j_n\ldots j_1}(\theta^\prime_n\ldots\theta^\prime_1) Y_{j_n\ldots j_1}(\theta_n\ldots\theta_1)\equiv \d(\Omega-\Omega^\prime)
\ee
where by definition
\be
\int d\Omega^\prime
\d(\Omega-\Omega^\prime)f(\theta^\prime)=f(\theta)
\ee 
whence in a somewhat symbolic form,
\be
\d(\Omega-\Omega^\prime)= \d(\theta^\prime_1-\theta_1)\ldots \d(\theta^\prime_n-\theta_n) \sin^{-(n-1)}\theta^\prime_n\,\sin^{-(n-2)}\,\theta^\prime_{n-1}\ldots \sin^{-1}\,\theta^\prime_2
\ee

\par
Now we can expand this function, as any other function, in series of Gegenbauer polynomials
\be\label{delta}
\d(\Omega-\Omega^\prime)=\sum_j d_j C^\n_j(\cos\,\theta_n)
\ee

Let us choose our reference frame in such a way that
\be
\Omega\cdot\Omega^\prime\equiv \cos\,\theta_n
\ee
id est, $\Omega^\prime$ is pointing towards the North pole.
\par
On functions constant on $S_{n-1}$,
\be
d\Omega_n=\frac{2\pi^{\frac{n}{2}}}{\Gamma(\frac{n}{2})}\sin^{n-1}\,\theta_n\,d\theta_n
\ee
and, denoting $x\equiv \cos\,\theta_n$
\be
d\Omega_n=\frac{2\pi^{\frac{n}{2}}}{\Gamma(\frac{n}{2})}\left(1-x^2\right)^{\frac{n-2}{2}} dx
\ee
as well as
\be
\d(\Omega)=\frac{\Gamma(\frac{n}{2})}{2\pi^{\frac{n}{2}}}\d(\theta_n)\frac{1}{\sin^{n-1}\,\theta_n}=\frac{\Gamma(\frac{n}{2})}{2\pi^{\frac{n}{2}}}\d(1-x)(1-x^2)^{\frac{2-n}{2}}
\ee
We can now integrate the two sides of the equation (\ref{delta}) against $C^\n_{j^\prime}(x)(1-x)^{\n- 1/2}$.
The orthogonality property 
\be
\int_{-1}^1 dx C^\n_j(x) C^\n_{j^\prime}(x)(1-x^2)^{\n-1/2}=\d_{j j^\prime}\frac{2^{1-2\n}\pi \Gamma(j+2\n)}{j! (\n+j)\Gamma(\n)^2}
\ee
then implies
\be
d_{j }\frac{2^{1-2\n}\pi \Gamma(j+2\n)}{j! (\n+j)\Gamma(\n)^2}=\frac{\Gamma(\frac{n}{2})}{2\pi^{\frac{n}{2}}}\int_{-1}^1 dx C^\n_j(x) (1-x^2)^{1 - n/2} \d(x-1)(1-x^2)^{\n-1/2}
\ee
The member of the right converges when $\nu=\frac{n-1}{2}$. Given in addition the fact that
\be
C_j^\n(1)=\frac{\Gamma(j+2\n)}{j!\, \Gamma(2\n)}
\ee
we can write 
\be
d_j=\frac{\Gamma(\frac{n}{2})(j+ \frac{n-1}2)\Gamma(\frac{n-1}{2})^2}{\Gamma(n-1) \pi^{\frac{n+1}2} 2^{3-n}}=\frac1{V(S_n)}\frac{n-1+2j}{n-1}
\ee

(using $\Gamma(2x)=2^{1-2x}\sqrt{\pi}\Gamma(x+\frac12)/\Gamma(x)$) as well as 
\be
\d(\Omega-\Omega^\prime)=\sum_j \frac1{V(S_n)}\frac{n-1+2j}{n-1}
C^{\frac{n-1}2}_j(\cos\,\theta_n)
\ee

\be
\sum_{j_n\ldots j_1} Y^*_{j_n\ldots j_1}(\theta^\prime_n=0\ldots\theta^\prime_1) Y_{j_n\ldots j_1}(\theta_n\ldots\theta_1)=\sum_j \frac1{V(S_n)}\frac{n-1+2j}{n-1}
C^{\frac{n-1}2}_j(\cos\,\theta_n)
\ee
If we employ the notation
$j\equiv j_n$ and $\vec{m}\equiv \left(j_{n-1}\ldots j_1\right)$, then the preceding formula presumably means that

\be
\sum_{\vec{m}} Y^*_{j\ldots \vec{m}}(\Omega_z) Y_{j\ldots \vec{m}}(\Omega)= \frac1{V(S_n)}\frac{n-1+2j}{n-1}
C^{\frac{n-1}2}_j(\cos\,\theta_n)
\ee

\item
We begin by defining some eigenfunctions of the differential operator:
\be
D\equiv\frac{\pd^2}{\pd\theta^2}+(N-1)\cot\,\theta\frac{\pd}{\pd\theta}-\frac{j\left(j+N-2\right)}{\sin^2\,\theta}
\ee
such that
\be\label{basica}
D \bar{P}^j_{Nk}(\theta)=-k\left(k+N-1\right)\bar{P}^j_{Nk}(\theta)
\ee
The form we are going to need is
\be
\left(\frac{\pd^2}{\pd\theta^2}+(N-1)\cot\,\theta\frac{\pd}{\pd\theta}\right)\bar{P}^j_{Nk}(\theta)=\left(\frac{j\left(j+N-2\right)}{\sin^2\,\theta}-k\left(k+N-1\right)\right)\bar{P}^j_{Nk}(\theta)
\ee
To be specific,

\be
\bar{P}^j_{Nk}(\theta)\equiv c^j_{Nk}\left(\sin\,\theta\right)^{-\frac{N-2}{2}} P^{-\left(j+\frac{N-2}{2}\right)}_{k+\frac{N-2}{2}}(\cos\,\theta)
\ee
where $P^\m_\n(z)$ are Legendre functions , and the normalization is given by
\be
c^j_{Nk}\equiv \sqrt{\frac{2k+N-1}{2}\frac{\left(k+j+N-2\right)!}{\left(k-j\right)!}}
\ee

The differential equation that Legendre functions $P^\m_\n\left(z\right)$ are solutions of is given by
\be
L w(z)\equiv \left(1-z^2\right)\frac{d^2 w}{dz^2}-2 z \frac{dw}{dz}+\left(\n\left(\n+1\right)-\frac{\m^2}{1-z^2}\right)w=0
\ee
Changing variables $z=\cos\theta$ this reads
\be
\left(\frac{\pd^2}{\pd\theta^2}+\cot\,\theta\frac{\pd}{\pd \theta}-\frac{\m^2}{\sin^2\,\theta}\right)w\left(\cos\,\theta\right)
=-\n\left(\n+1\right)w\left(\cos\,\theta\right)
\ee
and using this it is not difficult to actually prove the basic equation (\ref{basica}).
\par
The harmonics themselves are given by:
\be
Y_{j_n\ldots j_1}(\theta_n,\ldots, \theta_1)\equiv \prod _{m=2}^n \bar{P}_{m\, j_m}^{j_{m-1}}\left(\theta_m\right)\frac{1}{\sqrt{2\pi}}e^{i j_1\theta_1}
\ee
It is actually easy to check. From the expression for the laplacian, the operator acting on $\theta_1$, 
just leads to
\be
-\frac{j_1^2}{\sin^2\theta_n\ldots \sin^2\theta_2}
\ee
Next, the operator acting on $\theta_2$, corresponding to $N=2$,$k=j_2$ and $j=j_1$, yields

\be
\frac{j_1^2}{\sin^2\theta_n\ldots \sin^2\theta_2}-\frac{j_2(j_2+1)}{\sin^2\theta_n\ldots \sin^2\theta_3}
\ee
Next, the operator acting on $\theta_3$,
which corresponds to $N=3$, $k=j_3$ and $j=j_2$, gives
\be
\frac{j_2(j_2+1)}{\sin^2\theta_n\ldots \sin^2\theta_3}-\frac{j_3(j_3+2)}{\sin^2\theta_n\ldots \sin^2\theta_4}
\ee

After all pairwise cancellations, we are left with the last term, corresponding to $N=n$, $k=j_n$ and $j=j_{n-1}$, yielding the eigenvalue
\be
-j_n(j_n+n-1)
\ee

\item We can now employ the expansion (GR, 8.534)
\be
e^{im\rho \cos\,\phi}=2^\n \Gamma(\n)\sum_{k=0}^\infty (\n+k) i^k (m\rho)^{-\n} J_{\n+k}(m\rho) C^\n_k (\cos\,\phi)
\ee
and using our expansion of the Gegenbauer polynomials in terms of spherical harmonics,
\bea
&&e^{iz \Omega.\Omega^\prime}=2^{n/2 -1}\Gamma(n/2 -1)\sum_{k=0}^\infty (n/2 -1+k) i^k (z)^{-(n/2 -1)} J_{n/2 -1 +k}(z) \nonumber\\
&&C_{k,n}\sum_{\vec{m}} Y^*_{k\ldots \vec{m}}(\Omega) Y_{k\ldots \vec{m}}(\Omega^\prime)
\eea
where $C_{l,n}$ are apropiate constants.
\ei
\newpage

\newpage 

\end{document}